\documentclass[fleqn,usenatbib,usedcolumn]{mnras}
\usepackage[british]{babel}

\usepackage{mathptmx}

\usepackage[T1]{fontenc}
\usepackage{ae,aecompl}

\usepackage{graphicx}	
\usepackage{amsmath}	
\usepackage{amssymb}	
\usepackage{wasysym}

\DeclareSymbolFont{matha}{OML}{txmi}{m}{it}
\DeclareMathSymbol{\varv}{\mathord}{matha}{118}
\newcommand{\euler}{\text{e}}
\newcommand{\Mdot}{ \dot{M}}
\newcommand{\Msol}{ M_{\astrosun} }
\newcommand{\Rsol}{ R_{\astrosun} }
\newcommand{\kms}{km\,${\rm s}^{-1}$\,}
\newcommand{\yr}{\,\text{yr}^{-1}}

\title[Identifying  the companion of WR 148]{WR 148:  Identifying  the companion of an extreme runaway massive binary\thanks{Part of the data presented herein were obtained at the W.M. Keck Observatory, which is operated as a scientific partnership among the California Institute of Technology, the University of California and the National Aeronautics and Space Administration. The Observatory was made possible by the generous financial support of the W.M. Keck Foundation. }}

\author[M. Munoz et al.]{
Melissa Munoz,$^{1}$\thanks{E-mail: munoz@astro.umontreal.ca }
Anthony  F.J. Moffat,$^{1}$
Grant M. Hill,$^{2}$
Tomer Shenar,$^{3}$
\newauthor
Noel D. Richardson$,^{4}$
Herbert Pablo,$^{1}$
Nicole St-Louis$^{1}$
and Tahina Ramiaramanantsoa$^{1}$
\\
$^{1}$D\'{e}partement de Physique, Universit\'{e} de Montr\'{e}al, and Centre de Recherche en Astrophysique du Quebec, CP 6128, Succursale, Montr\'{e}al, \\ QC H3C 3J7, Canada\\
$^{2}$W.M. Keck Observatory, 65-1120 Mamalohoa, Kamuela, HI 96743, USA\\
$^{3}$Institut f\"urr Physik und Astronomie, Universit\"{a}t Potsdam, Karl-Liebknecht-Str. 24/25, 14476, Potsdam, Germany\\
$^{4}$Ritter Observatory, Department of Physics and Astronomy, The University of Toledo, Toledo, OH 43606-3390, USA\\
}

\date{Accepted. Received; in original form}

\pubyear{2016}

\volume{{\rm in press}}

\begin{document}
\label{firstpage}
\maketitle

\begin{abstract} 
WR 148 (HD 197406) is an extreme runaway system considered to be a potential candidate for a short-period (4.3173 d) rare WR + compact object binary. Provided with new high resolution, high signal-to-noise spectra from the Keck observatory, we determine the orbital parameters for both the primary WR and the secondary, yielding respective projected orbital velocity amplitudes of $88.1\pm3.8$\,\kms and $79.2\pm3.1$\,\kms and implying a mass ratio of $1.1\pm0.1$. We then apply the shift-and-add technique to disentangle the spectra and obtain spectra compatible with a WN7ha and an O4-6 star. Considering an orbital inclination of $\sim67^\circ$, derived from previous polarimetry observations, the system's total mass would be a mere 2-3 Msol, an unprecedented result for a putative massive binary system. However, a system comprising a $37\Msol$ secondary (typical mass of an O5V star) and a $33\Msol$ primary (given the mass ratio) would  infer an inclination of $\sim18^\circ$. We therefore reconsider the previous methods of deriving the orbital inclination based on time-dependent polarimetry and photometry. While the polarimetric results are  inconclusive requiring better data, the photometric results favour low inclinations. Finally, we compute WR 148's space velocity and retrace the runaway's trajectory  back to the Galactic plane (GP). With an ejection velocity of $198\pm27$\,\kms and a travel time of $4.7\pm0.8$\,Myr to reach its current location, WR 148 was most likely ejected via dynamical interactions in a young cluster. 
\end{abstract}




\section{Introduction}
Current massive-star evolution theories agree that Wolf-Rayet (WR) stars originate from O-type main sequence stars \citep{Crowther}. Thus, we expect a commonplace O+O binary system to evolve into an O+cc (compact companion), i.e. a high mass X-ray binary (HMXRB) if they remain bound after the first supernova explosion (SN). Recent population synthesis models suggest that roughly 80-95\,\% of the O+O binaries are disrupted subsequent to the SN \citep{Eldridge,lrr}. The surviving O+cc systems are then expected to progess to a WR+cc system. Yet, while the number of known O + cc sytems is significant, the number of detected WR+cc systems is much lower than predicted. In fact, while there are a total of 114 confirmed HMXRBs \citep{HMXRB} in our Galaxy, there is only a single confirmed WR+cc binary, Cyg X-3 \citep{CygX3}, compared to the $\sim10$ expected based on the Galactic fraction of WR- to O- stars \citep{VH7}.

HD 197406 (WR 148) is a well known candidate for such WR+cc systems. It is a single-lined spectroscopic binary with a well established $\sim4.3173$ d orbital period where the unseen companion was suspected to be either a low mass B2-B5 V-III star or a black hole (BH) \citep{Marchenko}. Found at roughly 800 pc above the Galactic plane \citep{Rosslowe}, HD 197406 is an extreme runaway with a high peculiar velocity. The compact companion is therefore an interesting possibility, as the recoil of the supernova explosion could have ejected the binary from the Galactic plane without breaking it up. The major shortfall of this hypothesis is the presence of a thermal X-ray spectrum \citep{XMM}, typical for WR+O colliding wind binaires, rather than a hard spectrum generally observed in accreting massive X-ray binaries.

The purpose of this investigation is therefore to determine the nature of the secondary in WR 148 and understand the evolutionary path of WR 148 as a unique runaway WR binary. In order to do so, we have acquired high resolution, very high signal-to-noise spectra at the Keck Observatory at both quadratures, complemented with a dozen lower quality spectra at the Observatoire du Mont M\'{e}gantic (OMM) during the summers of 2014 and 2015. The OMM data will serve mostly to refine the system's orbital parameters based on the bright primary WR component, whereas the Keck data will be used in an attempt to determine the nature of the companion.

The rest of the paper is organized as follows: We begin by describing the details of the observations in section \ref{Data}. This is followed by a thorough analysis of the observational data in section \ref{Analysis}. Finally, in section \ref{Conclusion}, we briefly summarize the results.

\section{Observational Data}
\label{Data}
Optical spectra were obtained during the summers of 2014 and 2015 at the Keck observatory and in 2014 at OMM. Table \ref{tab:obs} summarizes  the details of the observations.

A total of 36 spectra was obtained over the course of a month with the Perkin-Elmer optical long slit spectrograph mounted at the Cassegrain focus of the 1.6-m telescope of OMM. Each spectrum was extracted and reduced individually using standard IRAF\footnote{IRAF is distributed by the National Optical Astronomy Observatories, which are operated by the Association of Universities for Research in Astronomy, Inc., under cooperative agreement with the National Science Foundation} techniques. If multiple exposures were taken during one night, they were combined in one or more multiples to ensure a minimal spread in phase of 10\,\%. Once combined, we acquired 13 spectra, each providing a signal-to-noise ratio (S/N) of $\simeq$ 250 in the continuum.  

WR 148 was observed for two nights at the Keck observatory with the Echellette Spectrograph and Imager (ESI) instrument at roughly opposite quadratures: phase $\simeq$ 0.28 in the night of 2014 and phase $\simeq$ 0.73, a year later, in 2015. At each quadrature, all spectra were combined to produce one high S/N spectrum well above 1000 per pixel in the continuum.  

The data were rectified by selecting obvious regions of continuum and fitting a low order smooth function. For the echelle data, the overlaps in the orders were combined via a linear interpolation between pixels.

\begin{table*}
 \caption{Summary of spectroscopic observations.}
  \label{tab:obs}
\footnotesize
\begin{center}
  \begin{tabular}{ l c c c c c c c c}
    \hline \hline
     Telescope & Dates &  & No. of spectra & Exp. Time & Spectral range & Resolving power & S/N \\
           &  & & & [sec] & [\AA{}] &  & (combined$^{\dagger}$) & \\ \hline  
    OMM (1.6 m) & 2014 & July 25 - August 23 & 36 & 1800 & 3800 - 5000 & $\sim$3000 & $\sim$250 \\
    Keck (10 m) & 2014 & July 24 & 49 & 120 & 3900 - 6900  & $\sim$8000 & $\sim$1000 \\
    Keck (10 m) & 2015 & July 15 & 42 & 120 & 3900 - 6900 & $\sim$8000 & $\sim$100 \\ 
    \hline
  \end{tabular}
   ${\,}^{\dagger}$combined in 13 groups for OMM, while all spectra were combined for each of the Keck dates. 
\end{center}
\end{table*}

\section{Analysis and Results}
\label{Analysis}
\subsection{Orbital Elements}
\label{orbit}
We measure the radial velocities (RVs) of the N {\sc iv} $\lambda$4058 line by means of the bisector method (i.e. the RV refers to the centroid of the line). N {\sc iv} $\lambda$4058 is a relatively strong, narrow and symmetric line that is formed quite deep in the WR wind and is thus suspected to best reflect the true orbital motion of the WR. Though it is not uncommom for a late WN (WNL) star to have its  zero-point shifted by $\sim10$\kms to negative values  \citep{Moffat1979}. The quality of the OMM data alone, when taken comparatively from one spectrum to another, is not sufficient to detect obvious signatures of the secondary. Therefore, the companion lines should not affect the RV measurements for the WR orbit from most lines, but especially from N {\sc iv} $\lambda$4058. Applying the period finding algorithm described in \citet{SBMC}, capable of processing multiple datasets with large gaps in time simultaneously, to the present measurements in addition to all available anterior spectroscopic observations allows us to improve WR 148's orbital solution. The prevous RV measurrements are provided from \citet{Bracher}, \citet{Moffat1979}, \citet{Drissen} and \citet{Marchenko}. As noted in \citet{Marchenko}, in spite of the different methods used for the older determinations (e.g. fitting a parabola to the peak of the N {\sc iv} $\lambda$4058 line or bisector method), the RVs are in accordance and no signifiant consequences are reported.

\begin{figure}
    \centering
   \includegraphics[scale=0.4]{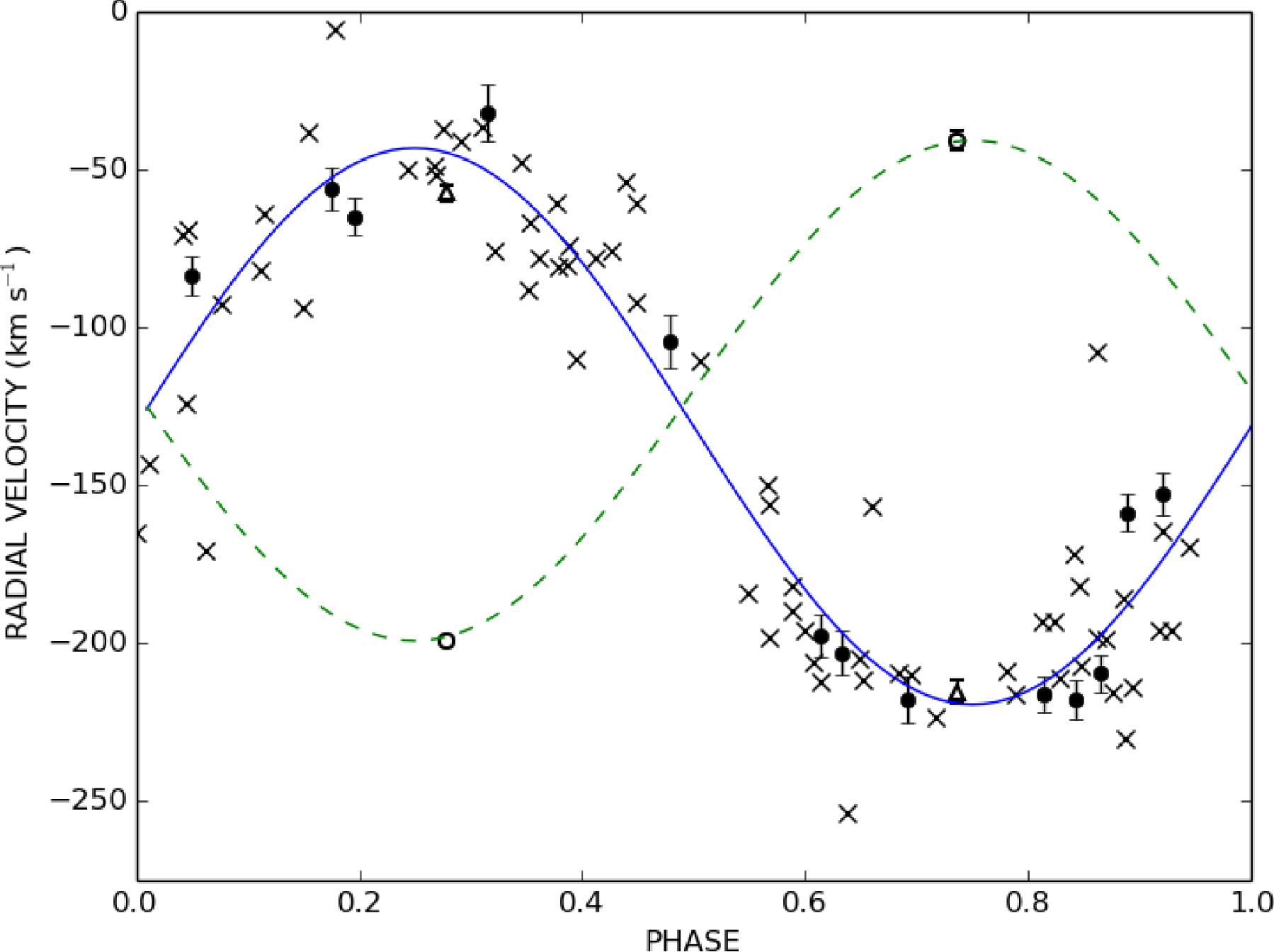}
    \caption{Radial velocity curve for the WR-star (blue solid line, based upon the N IV $\lambda$4058 line) and the O-type companion (green dashed line, based upon the He {\sc ii} $\lambda$4542, C {\sc iv} $\lambda$$\lambda$5801, 5812 and O {\sc iii}  $\lambda$5592. lines ). The crosses correspond to all previous observations from \citet{Bracher}, \citet{Moffat1979}, \citet{Drissen} and \citet{Marchenko} and the filled circles to the most recent 2014 observations from OMM. The hollow triangles and circles correspond to the Keck measurement for the WR and companion, respectively.  }
    \label{fig:RV}
\end{figure}

An elliptical orbit fit was first attempted followed by a circular fit (see Table \ref{tab:RV}). We note that the two sets of orbital elements have nearly the same quality of fit, as seen in the dispersion of the measured velocities of both fits being roughly identical: 24.10\kms for the elliptical fit and 24.06\kms for the circular fit. In a situation like this with a low eccentricity, the simple $p_1$ test described by Lucy (2005) quickly shows that the circular solution is strongly favoured. With the circular fit, we confirm and refine the previously reported period of P = 4.317336 d and time for phase zero (WR in front at inferior conjunction) of E = 2 444 825.04 HJD. The remaining orbital parameters are listed in Table \ref{tab:RV} and Fig. \ref{fig:RV} displays the WR RV-curve. The refined orbtital solution is in agreement with the most recent analysis from \citet{Marchenko}.


\begin{table}
\begin{center}
 \caption{Radial velocities for the WR component based on N {\sc iv} $\lambda$4058 from OMM and Keck data.}
  \label{tab:rv}
\footnotesize
  \begin{tabular}{ l c c }
    \hline \hline
     HJD - & Orbital phase & RV [\kms]  \\
     2 450 000		& 	 &	N {\sc iv} $\lambda$4058 	\\	 \hline 
6862.9773${\,}^{\dagger}$  &  0.2790  &   -57.6   $\pm$    2.6	\\     
 6864.7613  &   0.6922   &  -217.9   $\pm$      7.2	\\
 6865.6098  &   0.8887   &  -158.7   $\pm$      6.0	\\
 6865.7480  &   0.9208   &  -152.7   $\pm$      6.7	\\
 6868.7376  &   0.6132   &  -197.5   $\pm$      6.7	\\
 6868.8226  &   0.6329   &  -203.1   $\pm$      6.9	\\
 6869.6056  &   0.8143   &  -216.4   $\pm$      5.6	\\
 6869.7292  &   0.8429   &  -218.0   $\pm$      6.2	\\
 6869.8249  &   0.8651   &  -209.6  $\pm$     6.0	\\
 6870.6204  &   0.0493   &   -83.9   $\pm$      6.1	\\
 6871.7661  &   0.3147   &   -31.9   $\pm$      9.0	\\
 6889.7406  &   0.4780   &   -104.7   $\pm$      8.4	\\
 6892.7523  &   0.1756   &   -56.4   $\pm$     6.7	\\
 6892.8417  &   0.1963   &   -65.2   $\pm$      5.9 \\
7218.9720${\,}^{\dagger}$  &  0.7360   &  -215.4   $\pm$    3.4 \\
    \hline
  \end{tabular}
      \\ ${\,}^{\dagger}$RV measurements are from Keck Observatory.
\end{center}
\end{table}

\begin{table}
\begin{center}
 \caption{Orbital Elements for the WR component and derived quantities. $f(m_1)$ is the mass function for the WR-star given by $f(m_1) \equiv P K_1^3/2 \pi G = (m_2 \sin i)^3/(m_1 + m_2)^2$.} 
  \label{tab:RV}
\footnotesize
  \begin{tabular}{ l c c}
    \hline \hline
     Parameter & Elliptical fit & Circular fit \\ \hline 
    $P$ [days] & 4.317340 $\pm$ 0.000026 & 4.317336 $\pm$ 0.000026 \\
    $T_0$ [HJD - 2,440,000] & 4818.5 $\pm$ 0.8 & - \\
    $E$ [HJD - 2,440,000]  & - & 4825.04  $\pm$    0.03 \\
    $e$					 & 0.05 $\pm$ 0.04 & 0 (fixed) \\
    $\omega$ [deg]		& 79.9 $\pm$ 64.7  & - \\
    $K_1$ [\kms] & 88.1 $\pm$    3.8 & 88.1 $\pm$    3.8 \\
    $\gamma_1$ [\kms]& -131.9 $\pm$ 2.7 & -131.4 $\pm$ 2.7 \\ 
    $a_1$ $\sin i$ $[R_{\astrosun}]$ & 7.5 $\pm$  0.3 & 7.5 $\pm$  0.3 \\   
    $f(m_1) $ $[M_{\astrosun}]$ & 0.30 $\pm$ 0.04  & 0.30 $\pm$ 0.04\\  
    \hline
   \end{tabular}
\end{center}
\end{table}

\begin{figure*}
    \centering
    \includegraphics[scale=0.4,trim={2cm 0 0 0},clip]{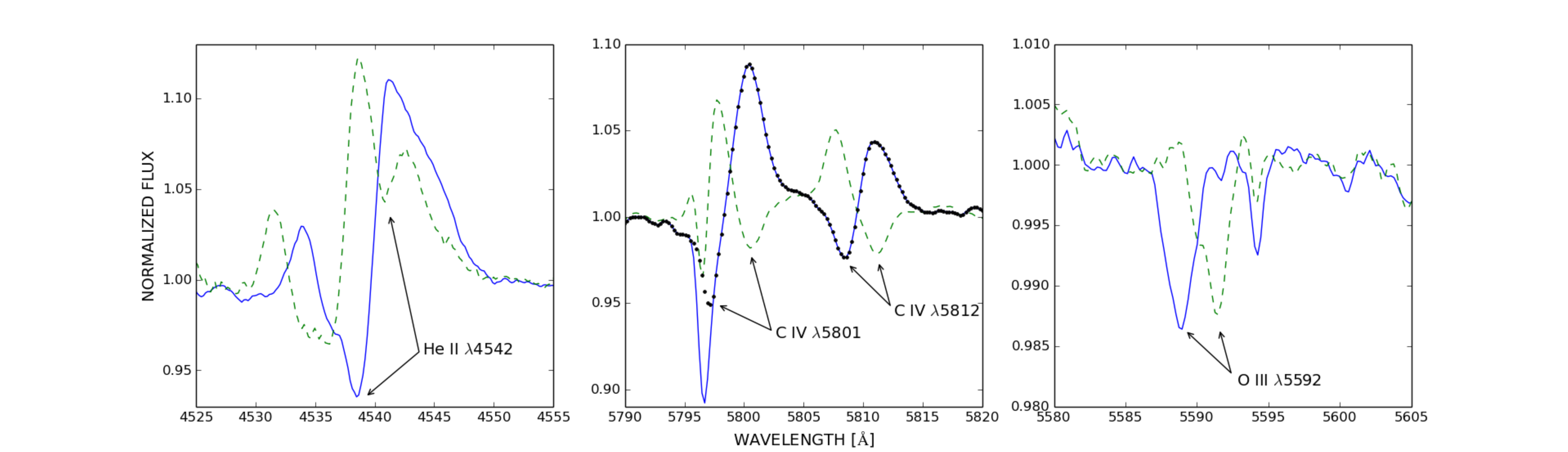}
    \caption{Most clearly revealed companion-star absorption lines from the Keck spectra, where measuring the RV was easily possible. The blue solid lines correspond to the first quadrature and the green dashed lines the second quadrature. The black dotted line is the attempted DIB-corrected spectrum for the first quadrature of the C {\sc iv} line pair. The companion's absorption lines are indicated with arrows and are, from panels left to right, He {\sc ii} $\lambda$4542, C {\sc iv} $\lambda$$\lambda$5801, 5812 and O {\sc iii}  $\lambda$5592.   }
      \label{fig:Oabs}
\end{figure*}

The Keck observations are of utmost importance for spectroscopically resolving this system. In fact, the higher resolution and higher S/N Keck data reveal the presence of absorption dips which move in clear anti-phase to the WR emission lines throughout the spectra in most of the H {\sc i}, He {\sc i} and He {\sc ii} lines.  We identify four absorption lines that were particularly isolated to characterize the orbit of the secondary: He {\sc ii} $\lambda$4542, C {\sc iv} $\lambda$$\lambda$5801, 5812 and O {\sc iii}  $\lambda$5592. Of these lines,  O {\sc iii} is the only absorption line unperturbed by WR emission lines. Most likely originating from the companion, we can deduce the secondary's gamma velocity, $\gamma_2$, and amplitude, $K_2$, from RV measurements at both epochs, assuming a circular orbit.  The RVs of the absorption lines were  derived by fitting a Gaussian to the trough of the line and are tabulated in Table \ref{tab:rvabs}. The resulting $\gamma_2$ and $K_2$ values are provided in Table \ref{tab:RVabs}. We note that, at the first quadrature, the C {\sc iv} $\lambda$5801 absorption line is slightly contaminated by a superimposed diffuse interstellar band (DIB) at $\lambda$5796.98 \citep{Herbig}.  An attempt was made to remove this DIB by subtracting a DIB profile obtained from the first qudrature to the second (see Fig. \ref{fig:Oabs}). However, we note an increase of DIB depth from the first quadrature to the second. This can seen at the DIB near $\sim5595$ \AA \citep{Herbig} in figure \ref{fig:Oabs}. Consequently, the correction will likely be underestimated.  In spite of this, the calculated $\gamma_2$ and $K_2$ velocities tie in well with respective values of $-120.1 \pm 1.2$\,\kms and $79.2 \pm 3.1$\,\kms.


\citet{Marchenko} did report the presecence of weak absorption features on top of H$\gamma$ $\lambda$4340 and He {\sc i} $\lambda$4471. Presumed to arise from the companion, \citet{Marchenko} obtained $K_1=87.7\pm2.4$\kms and $K_2=36.2\pm2.4$\kms.  While $K_1$ is compatible with our result, $K_2$ is not. However, the RV's for the companion were only seen and measured on 3 (for  H$\gamma$ $\lambda$4340) or 6 (for He {\sc iv} $\lambda$4471) of their 22 collected spectra. Given the sparsity of their measurents, the orbital solution may be unreliable. Futhermore, as the absorption dips are heavily blended with the WR emission lines, we suspect that some measurements could have been inaccurate.

\begin{table*}
\begin{center}
 \caption{Radial velocities for the companion based on He {\sc ii}  $\lambda$4542, C {\sc iv} $\lambda$5801,  C {\sc iv} $\lambda$5812 and O {\sc iii} $\lambda$5592 from Keck data (phases calcuated assuming period and time of phase zero from the WR component).}
  \label{tab:rvabs}
  \footnotesize
  \begin{tabular}{ l c c c c c  }
    \hline \hline
   HJD - & Orbital phase &    \multicolumn{4}{c}{RV [\kms]} \\ 
   2 450 000		& 	 & He {\sc ii}  $\lambda$4542 & C {\sc iv} $\lambda$5801 & C {\sc iv} $\lambda$5812 & O {\sc iii} $\lambda$5592	\\	 \hline 
  6862.9773  &   0.2790   &  -201.5 $\pm$ 2.6 & -212.9 $\pm$ 2.7 & -190.0 $\pm$ 3.0 & -191.3 $\pm$ 2.4\\
 7218.9720  &   0.7360   & -38.5 $\pm$ 4.0 &  -33.5 $\pm$ 2.5 & -41.7 $\pm$ 2.4 & -51.2 $\pm$ 1.5\\
      \hline 
  \end{tabular}
\end{center}
\end{table*}

We point out that the companion's  systemic velocity differs slightly from that of the WR-star.  Although the two values are marginally the same within the errors, we keep in mind that the WR's radial velocity is solely based upon the ability of the N {\sc iv} $\lambda$4058 line to reflect the true orbital motion of the WR component. However, because of the WR-star's substantial stellar wind there is some leeway, and therefore the companion's systemic velocity may be more reliable \citep{Moffat1979}. For this reason, we will adopt the companion's $\gamma_2$ as the true systemic velocity.

\begin{table}
\begin{center}
 \caption{Derived systemic velocities and amplitudes for the companion star from Table \ref{tab:rvabs}}
  \label{tab:RVabs}
  \begin{tabular}{ l c c  }
    \hline \hline
      Line &  K$_2$ [\kms] & $\gamma_2$ [\kms]  \\
      \hline
      He {\sc ii}  $\lambda$4542 & 82.3 $\pm$ 2.3 & -120.5 $\pm$ 2.3 \\
      C {\sc iv} $\lambda$5801 & 89.6 $\pm$ 2.5& -122.5 $\pm$ 2.3 \\
      C {\sc iv} $\lambda$5812 & 74.9 $\pm$ 2.8 & -116.3 $\pm$ 2.8  \\
      O {\sc iii} $\lambda$5592 & 72.0 $\pm$ 2.5  & -122.9 $\pm$ 2.5 \\
      \hline 
  \end{tabular}
\end{center}
\end{table}

Interestingly enough, the $K_1$ and $K_2$ values are quite similar, which would imply a mass ratio on the order of unity. As stated in \citet{Drissen}, an inclination angle of $\simeq 67^\circ$  was derived from polarimetric variations. This would infer a total mass of the system 
\begin{equation}
m_1 + m_2 = \frac{P }{2 \pi G} \frac{(K_1 + K_2)^3}{\sin^3 i} \simeq 2.69 \, \Msol,
\end{equation}
where subscript 1 refers to the WR-star and subscript 2 to the companion and $G$ is the gravitational constant. Either we are dealing with a low-mass [WR] central star of a planetary nebula (CSPN) or a nova-like V Sge star \citep{VSag} with a sub-dwarf OB companion, or the inclination angle is much lower than anticipated and the system is in fact a massive WR+OB binary. To resolve this situation, we attempt to rederive the orbital inclination via spectral classification of the secondary. With this new perspective, we review the previous polarimetric and photometric methods of determining $i$ in order to merge all the information.

\subsection{High or low mass?}

Could WR 148 be a low-mass WR-star, i.e.\ a central star of a planetary nebula or a  nova-like V Sge?
Although almost all known [WR] stars belong 
to the carbon sequence [WC], a few [WN] stars were found in recent years \citep{a,
b}, and
we therefore cannot reject the possibility of WR 148 being a rare case of a 
[WN] binary. While no nebular lines are present in the spectra at hand, this could 
be explained by the faintness of any surrounding nebula, or may even suggest a 
dissipation of the nebula due to binary interaction.

Unfortunately, [WR] (and V Sge) stars are virtually almost indistinguishable spectroscopically
from their massive counterparts. While mass-loss rates of [WR] stars are clearly 
smaller than those of massive WR-stars, 
so are their emitting surfaces, leading to comparable equivalent widths of the 
emission lines in both types of stars.
However, one somewhat subtle difference exists. 
Photons originating in recombination lines 
are scattered off free electrons in the WR wind and 
lead to the formation of so-called electron scattering wings (ESWs). ESWs 
form on both sides of the line (red and blue wings), 
but because the scattering
electrons move outwards with the expanding stellar wind, the red wing 
is more prominent than the blue wing \citep{c}.
The strengths of these wings are directly 
proportional to the mass-loss rate and 
to the strengths of the recombination lines from 
which they stem. 
Since mass-loss rates of [WR] stars are at least an order of magnitude 
smaller than those of massive WR-stars, they are expected
to show less pronounced or even vanishingly small ESWs.

\begin{figure}
    \centering
    \includegraphics[scale=0.4]{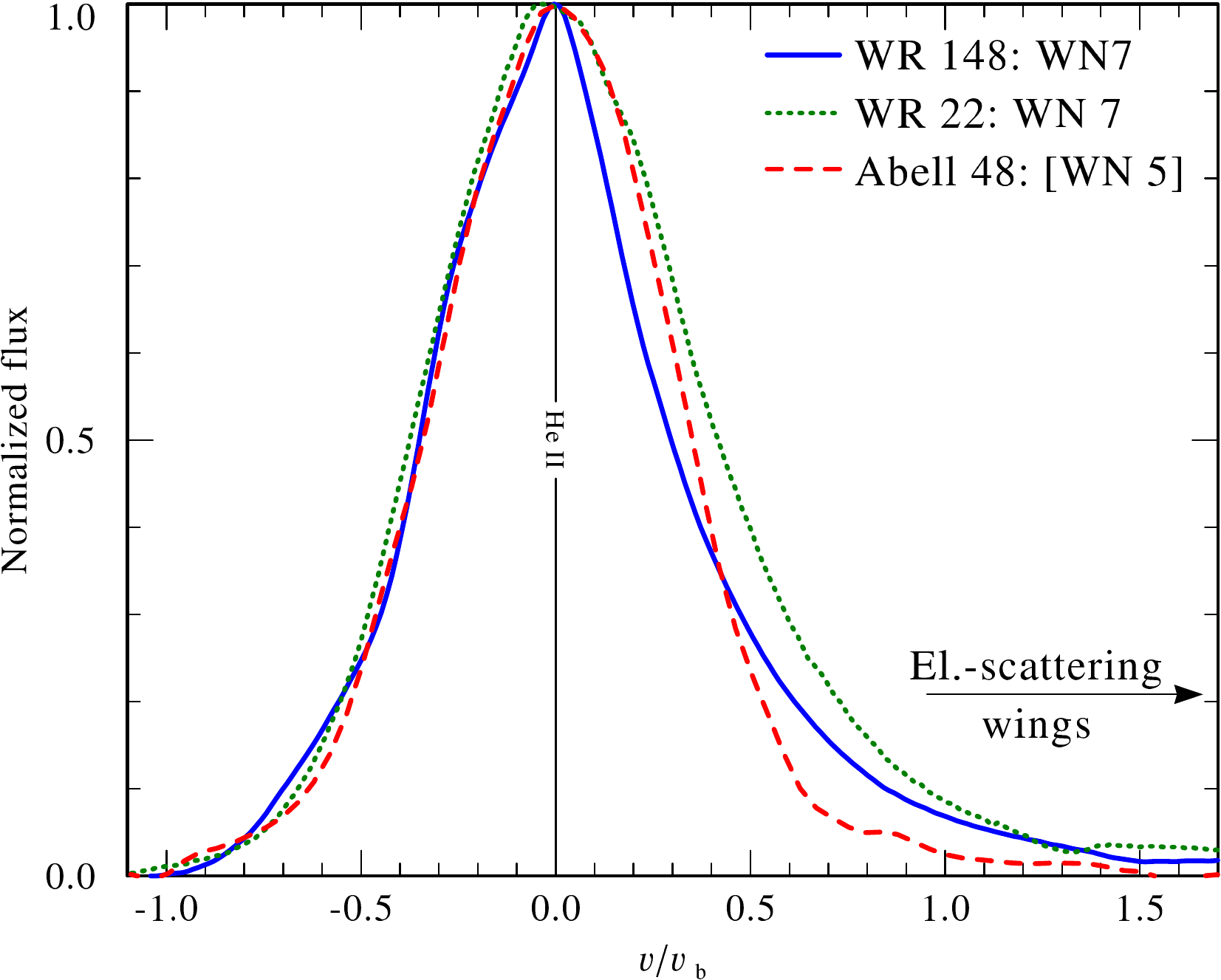}
    \caption{A comparison between the observed spectrum of WR 148 (blue solid line) 
    and a low-mass [WN] star, Abell 48 (red dashed line). The observation
    of Abell 48 was obtained with the South African Large Telescope (SALT), see 
    \citet{e} for further details. An observed spectrum of WR 22, a classical WN7 star, has also been added (dotted green line) for comparison purposes.}
    \label{fig:WRcomp}
\end{figure}

Fig. \ref{fig:WRcomp} shows a comparison between our observed spectra of WR 148, a WN7ha star (see section \ref{Class}), and 
Abell 48, a [WN5] central star  \citep[see][]{e}, focusing on the strong He\,{\sc ii}
$\lambda 4686$ 
line. Identifying the ESWs becomes easier when comparing the red and blue wings
of the line.
For each line, we measured the velocity corresponding to 
the blue edge of the line, i.e. where the profile reaches zero, $v_\text{b}$. We note that generally $v_\text{b}$ need
not be identical to $v_\infty$, 
since the width depends on the line-formation region and the amount of 
electron scattering present.
We find $v_\text{b} = 850$ and $1500\,$\kms 
for WR 148 and Abell 48, respectively.
The profiles shown in Fig. \ref{fig:WRcomp} are normalized to their respective peak intensities, 
and the wavelengths are transformed to the velocity space normalized to  
$v_\text{b}$. 

Note how the spectrum of WR 148 exhibits a clear asymmetry towards the red part of 
the line. While the emission goes to zero at $-v_\text{b}$ (per definition) on the blue side, there
is clearly emission 
excess beyond $v_\text{b}$ on the red side. This is a clear signature of the red ESW.
In contrast, the profile exhibited by 
Abell 48 appears rather symmetric, just as one would expect in the case of 
a central star given their low mass-loss rates in an absolute sense. Its ESWs are almost undetectable.
While the effect is subtle, it is apparent. 
Though the extent of the ESW will vary depending on the physical properties of the WR star, its presence should be noticeably larger in high-mass WR stars than low-mass [WR] stars, regardless of the ionization sequence.
We also compared the ESWs exhibited by WR 148 with other massive WR-stars and 
find them to be similar in strength. This gives further evidence that the WR
component of 
WR 148 is indeed a massive WR-star.

\subsection{Spectral Classification}
\label{Class}
\begin{figure*}
\begin{minipage}{\textwidth}
\includegraphics[scale=0.4,trim={2cm 0 0 0},clip]{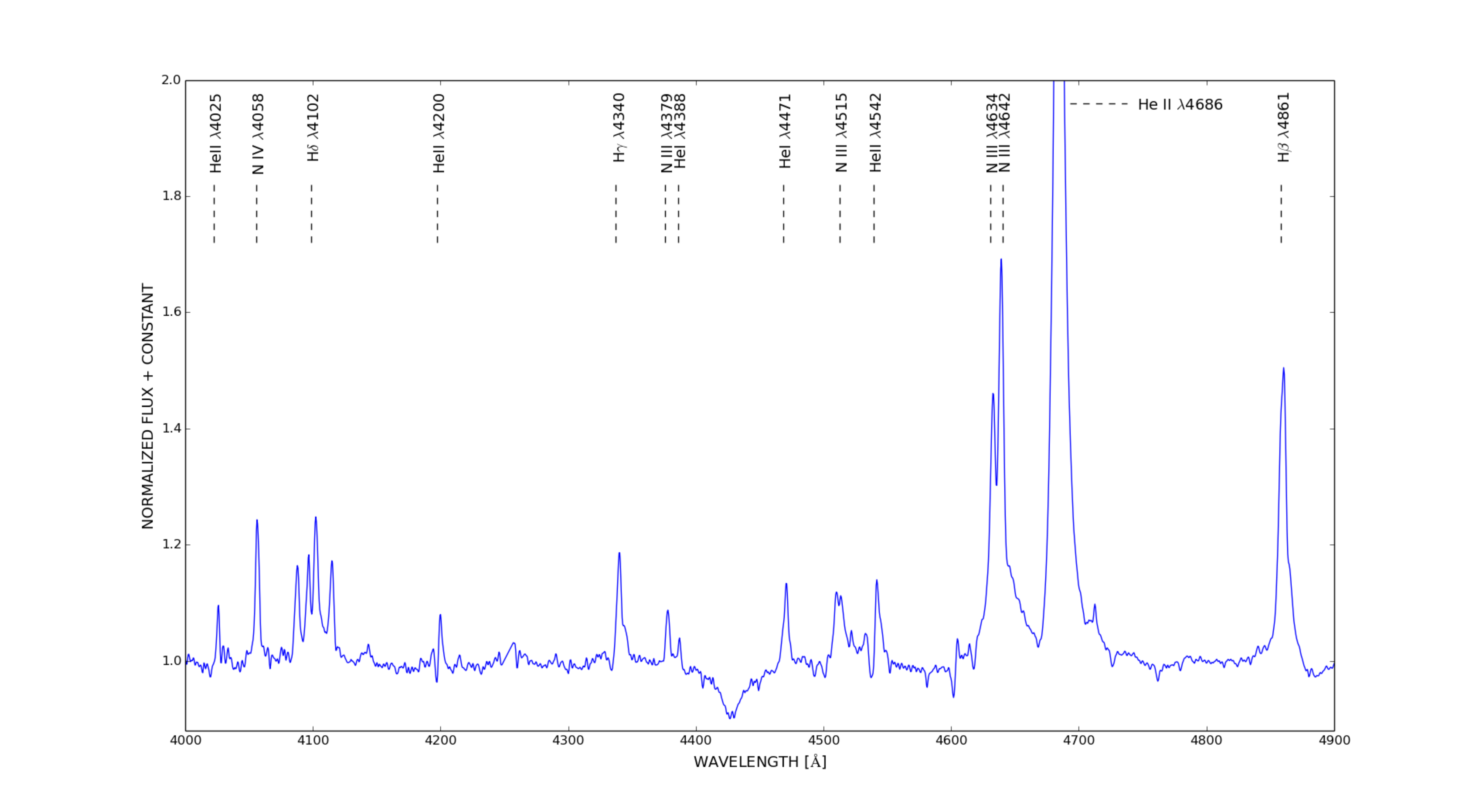}
\caption{Sample of WR 148's spectrum from one of the 13 OMM combined spectra.}
   \label{fig:WRomm} 
\end{minipage}
\begin{minipage}{\textwidth}
\includegraphics[scale=0.4,trim={2cm 0 0 0},clip]{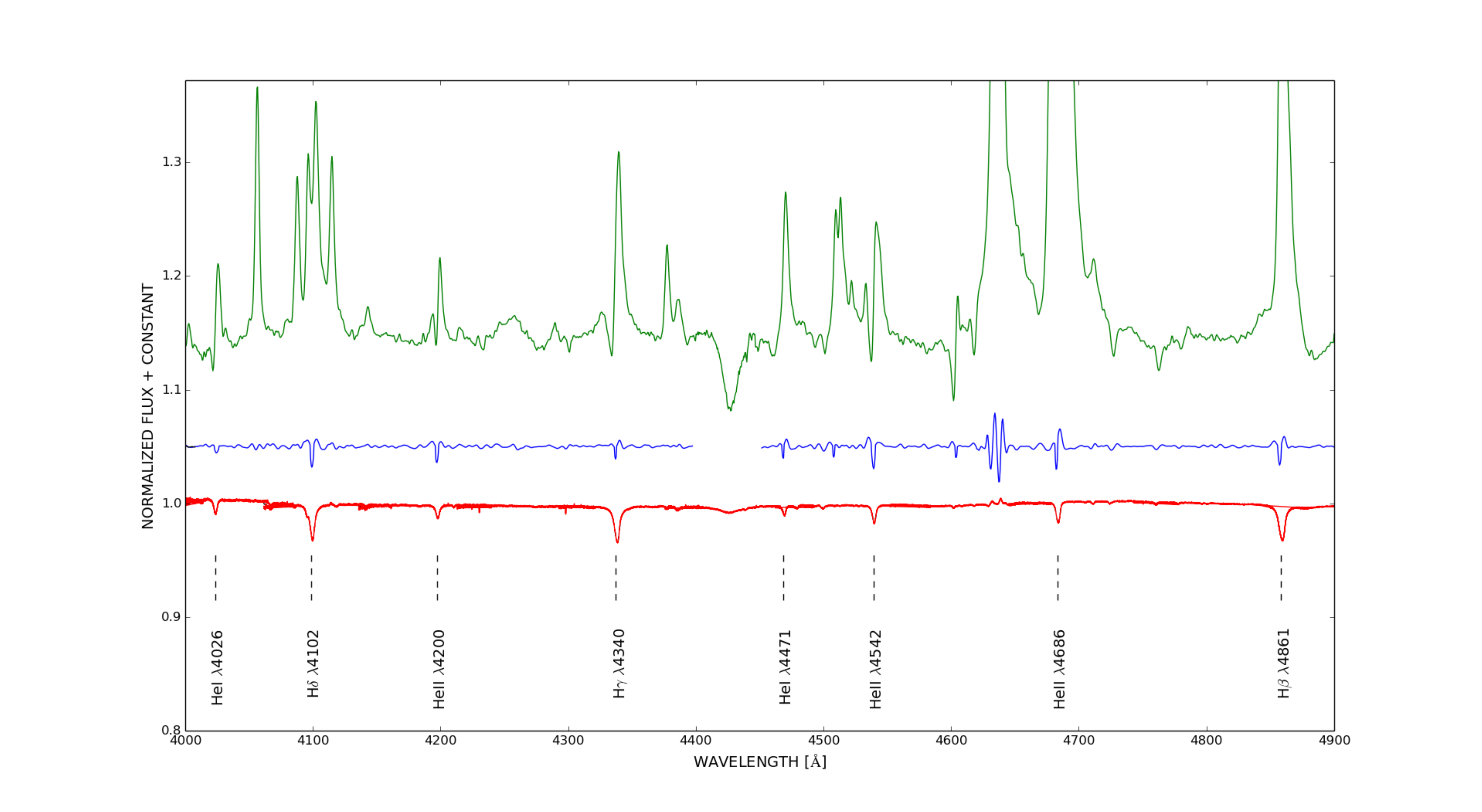}
\end{minipage}
\caption{Mean spectra of the WR 148 components after applying the shift-and-add separation. Above is the WN7ha (shifted vertically for clarity) and beneath, the presumed O5V spectrum. Below, a spectrum of an O5V standard star (HD 46150) is shown for comparison purposes \citep[provided from][]{PolarBase}.  Note that the absolute line strengths are not preserved in this procedure. }
   \label{fig:shiftadd}
\end{figure*}

WR 148 was re-classified as a WN8h in \citet{Smith} according to 3D their classification guide. However, we reaffirm the former WN7 spectral type as quoted in \citet{Bracher}. This is based upon the line ratio of He{\sc{ii}} $\lambda$5411 to He{\sc{i}} $\lambda$5876 (Peak/Continuum) totalling $\simeq$ 0.66 for WR 148 (measured from the Keck spectra).  Furthermore, the H$^+$/He$^{++}$ fraction is well beyond the 0.5 threshold associated with WN7(h). Finally, with the added presence of intrinsic absorption lines we obtain WN7ha. A sample of the spectrum is shown in Fig. \ref{fig:WRomm}

The companion's spectral classification is based upon the ratio of the equivalent widths (EW) of He {\sc{i}} $\lambda$4471 to He {\sc{ii}} $\lambda$4542 \citep{Conti1977,STO}.  Since the OB absorption lines are blended with the WR emission lines, disentangling the spectra is crucial. We adopt the shift-and-add technique devised by \citet{ShiftAdd}. This procedure consists of first shifting all the spectra to the WR frame and creating a mean WR template. The latter is then subtracted from all the individual spectra and the result is shifted back to the O-star frame. Once combined, the output is the mean O-star spectrum. This is then substracted off the original spectra and the whole process is repeated as often as needed until a stable result is reached. Three iterations sufficed in this case.

This method requires knowledge of the orbital solution for both stellar components and warrants a dense enough data set with good phase coverage to smear out the companion's absorption lines. The Keck spectra supply the missing companion's orbital parameters, and are complementary to the OMM data, with evenly spaced phase coverage. With these  criteria fulfilled, we have extracted the mean O-star spectrum with only minor numerical artifacts on the companion's main He {\sc{i}} and He {\sc{ii}} lines (see Fig. \ref{fig:shiftadd}). Still, reconstructing the region between $\lambda$4600 - 4700 of the companion's spectrum was unsuccessful. In this region, the prominent features of the WR-star, (He {\sc{ii}} $\lambda$4686 and N {\sc{iii}} $\lambda$$\lambda$4634-4642 lines), are highly variable and overshadow the O-star absorption dips. Furthermore, a common residual side effect of this method is the presence of rounded emission-like edges on the absorption profiles. An accumulation of errors while subtracting the mean WR spectra from the individual spectra is most likely the cause.

Nonetheless, the He {\sc{i}} line strengths are noticeably weaker than for He {\sc{ii}}. Measuring the EW of (the absorption part of) He {\sc{i}} $\lambda$4471 and He {\sc{ii}} $\lambda$4542, yields $\log \left(  EW He \text{{ \sc{i} }} \lambda 4471/ EW He \text{{ \sc{ii} }}  \lambda4524 \right) = -0.51$. This corresponds to a spectral type of O5 for the companion \citep{STO}. For good  measure, we will consider the O4 to O6 range as a reasonable uncertainty window. As for the companion's luminosity classification, the weak relative line strength (peak to continuum) of the secondary to the WR star suggets the secondary is a dwarf. If the secondary were to be a giant or especially a supergiant, the relative line strength would be larger than the observed value of $\sim0.2$. We therefore presume the secondary is an O5V star.

\begin{figure}
    \centering
    \includegraphics[scale=0.4]{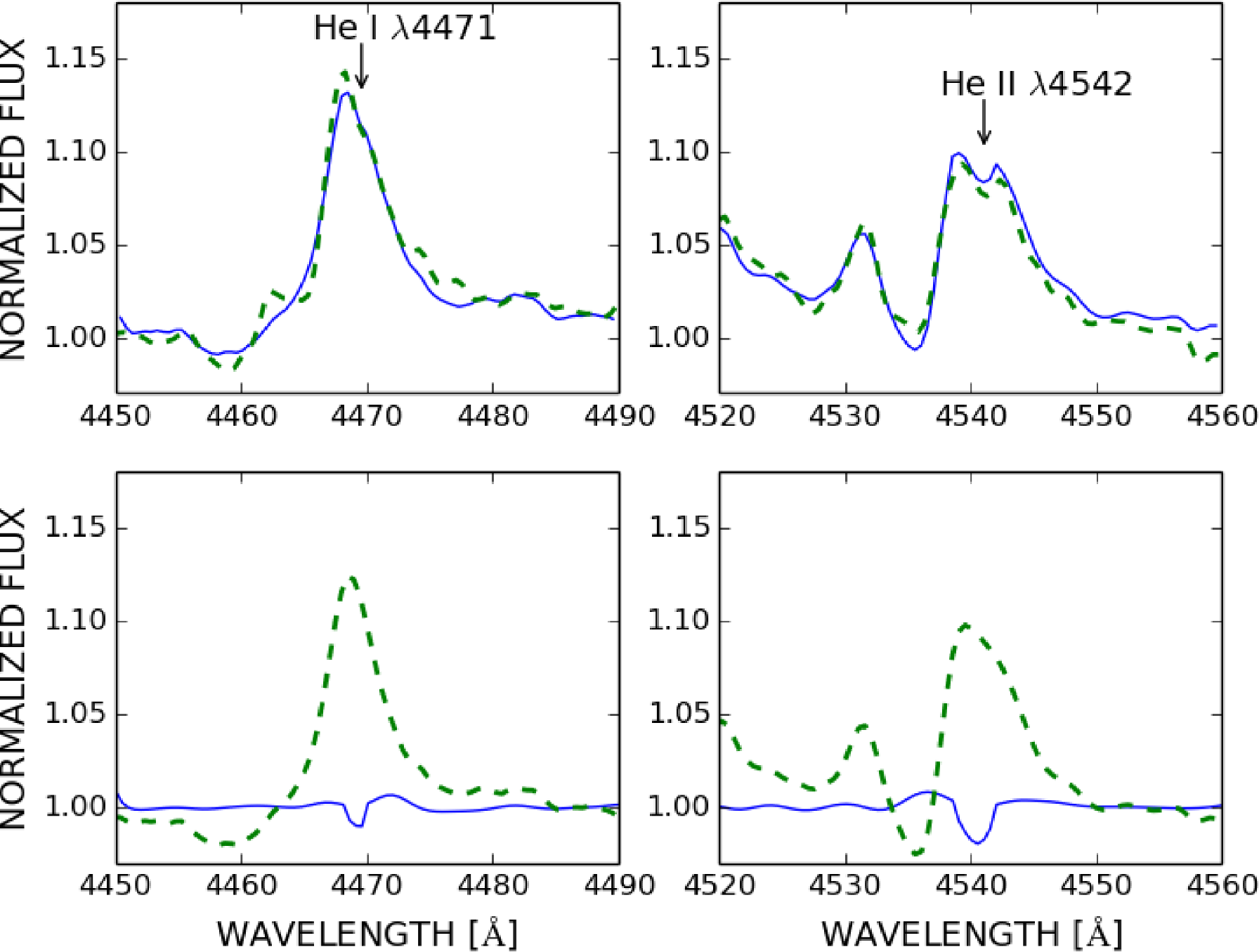}
    \caption{Top two panels: Comparison of the blended OMM spectra  (dashed curve) and the sum of the disentangled components from the top panels (solid curve) for He {\sc{i}} $\lambda$4471 (left panel) and He {\sc{ii}} $\lambda$4542 (right panel) at 0.61 orbital phase. Bottom two panels: The indivual components of the disentagled spectra for the WR- (dashed curve) and O-star (solid curved) zoomed on the He {\sc{i}} $\lambda$4471 (left panel) and He {\sc{ii}} $\lambda$4542 (right panel) lines.}
    \label{fig:new}
\end{figure}

We show in Fig.\ref{fig:new} the spectra of the disentangled components plotted over the blended spectra (from OMM at 0.61 phase) for both the He {\sc{i}} $\lambda$4471 and He {\sc{ii}} $\lambda$4542 lines. We can see that the superposition of the separated WR- and O-star spectra mimic well the composite spectrum. An improvement of the standard shift-and-add procedure from \cite{ShiftAdd} can be done following the method from \citet{ShiftAdd2} where after each iteration, the RVs for both components  are remeasured and consequently readjusted enabling one to refine the orbital solution. This is done at the end of each iteration by measuring the RVs from the initial composite spectrum after removal of one of the componenets. From this we obtain $K_2=78.5\pm4.7$ \kms and $\gamma_2=-124.9\pm4.4$\kms which is consistent with our previous results.

\begin{figure}
    \centering
    \includegraphics[scale=0.4]{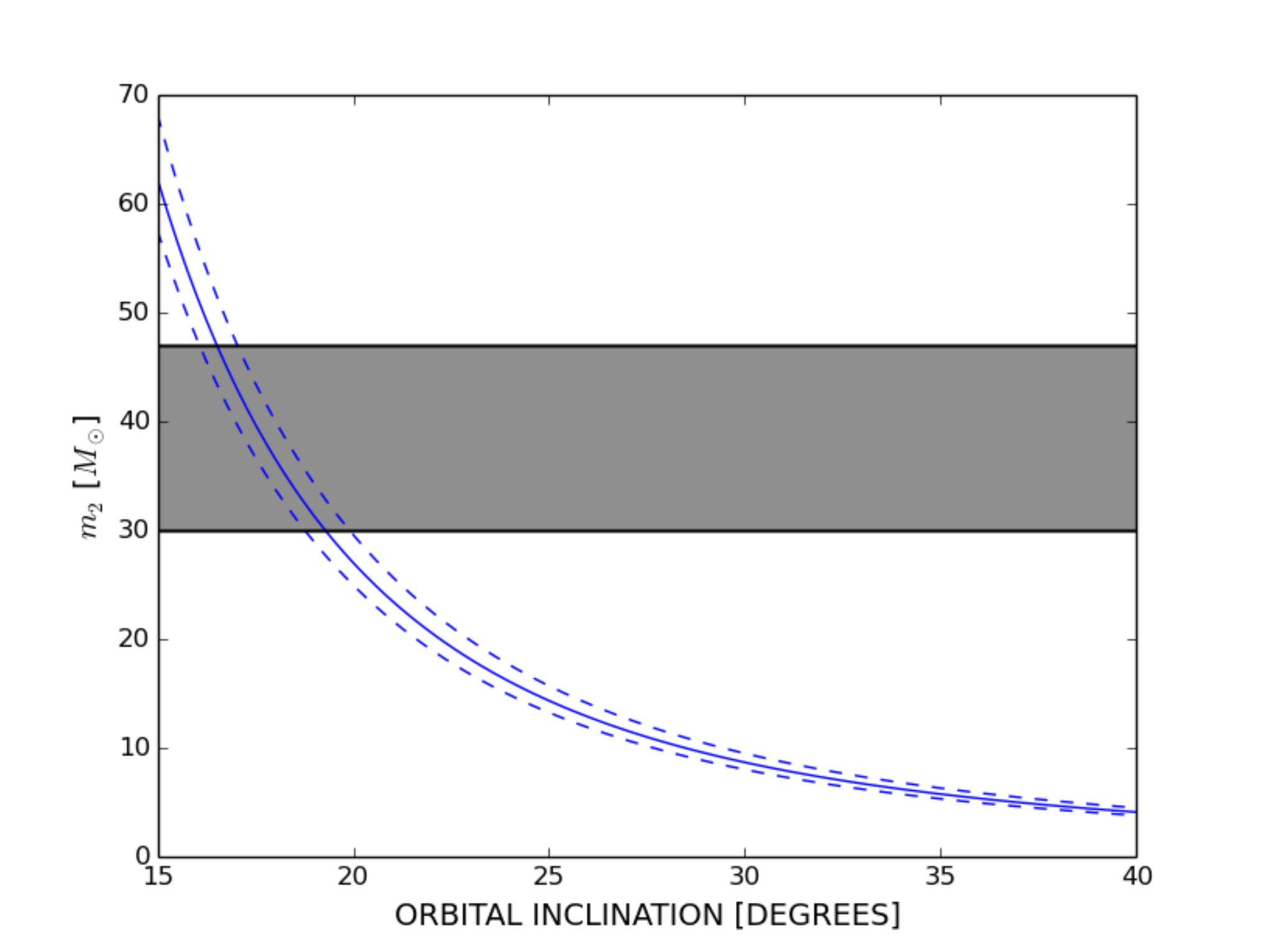}
    \caption{Mass of the companion as a function of orbital inclination for $q=1.1$ (blue solid line) $\pm0.1$ (blue dashed lines). The shaded area corresponds to possible inclination angles given the uncertainties on spectral types. The upper limit is the mass of a typical O4 star and lower limit for an O6 star \citep{ParamsO}}
    \label{fig:m2vsi}
\end{figure}

It is now possible to obtain the inclination angle via the mass function,
\begin{equation} \label{eq:fm}
f(m_1) \equiv \frac{P K_1^3}{2 \pi G} = \frac{(m_2 \sin i)^3}{(m_1 + m_2)^2}=0.30 \pm 0.04 \,\Msol,
\end{equation}
if the masses of both components are known. While the masses of WR-stars are not so well constrained based on their spectral types, the spectroscopic masses of OB-stars on the other hand are fairly well known as a function of spectral type. Knowing the mass ratio, $q=m_2/m_1=K_1/K_2 = 1.1 \pm 0.1$, we therefore only need to determine the mass of the companion. For an O5V star, $M_O =37\,\Msol$ \citep{ParamsO} and thus $i=18 \pm 4^\circ$. The uncertainty of $\pm$1 spectral class, has little affect on the inclination angle. This is because, at low inclinations, the values of the masses will vary drastically with only small changes of $i$ on account of $\sin i$ being to the third power in equation \ref{eq:fm} (see Fig. \ref{fig:m2vsi}). We assume zero error on the luminosity class.

\subsection{Revisiting the polarimetry}
\label{pol}
A linear polarization variability study on WR 148 was carried out by \citet{Drissen}. An inclination of $\simeq67^\circ$ was obtained by fitting separately the $q_i$ and $u_i$ Fourier coefficients to the following expressions for the $Q$ and $U$ Stokes parameters
\begin{equation}
  \begin{aligned}
Q&=q_0 + q_1 \cos(\lambda) + q_2 \sin (\lambda) +  q_3 \cos(2 \lambda) + q_4 \sin (2 \lambda), \\
U&=u_0 + u_1 \cos(\lambda) + u_2 \sin (\lambda)  + u_3 \cos(2 \lambda) + u_4 \sin (2 \lambda),
  \end{aligned}
\end{equation}
where $\lambda=2 \pi \phi$ and $\phi$ is the orbital phase with WR inferior conjunction at $\phi=0$.

This model can be further simplified assuming symmetry perpendicular to the orbital plane. Using the terminology from \citet*{BME}, hereafter BME, this signifies that $\tau_1=\tau_2=0$, where $\tau_i$ is the $i^\text{th}$ electron density moment. Under these circumstances, equations (3) re-write to 
\begin{equation}
  \label{eq:QU2}
  \begin{aligned}
    Q&= Q_0 + \Delta Q \cos \Omega - \Delta U \sin \Omega, \\
    U&= U_0 + \Delta Q \sin \Omega - \Delta U \cos \Omega,
  \end{aligned}
\end{equation}
with
\begin{equation}
  \begin{aligned}
\Delta Q &= -H\left[ (1+\cos^2 i)\cos 2(\lambda - \lambda_0 ) - \sin^2 i \right], \\
\Delta U &= - 2 H\cos i \sin 2 (\lambda - \lambda_0 ),
  \end{aligned}
\end{equation}
and
\begin{equation}
\label{eq:H}
 \begin{aligned}
H &= \sqrt{\tau_3^2 + \tau_4^2}, \\
\tau_3 &= H\cos 2 \lambda_0, \\
\tau_4 &= -H\sin 2 \lambda_0, 
  \end{aligned}
\end{equation}
where $Q_0$ and $U_0$ are interstellar values, $\Omega$ is the rotation angle of the line of nodes on the sky, $\tau_3$ and $\tau_4$ are the third and fourth electron density moments and $\lambda_0$ is a constant. 

By explicitly expressing the $q_i$ and $u_i$ coefficients as a function of the  independent parameters ($Q_0$,  $U_0$,   $i$, $\Omega$, $H$ and $\lambda_0$), it becomes apparent that the coefficients are in fact all coupled. Therefore, rather than fitting  $Q$ and $U$ with $q_i$ and $u_i$ as free parameters, we fit $Q$ and $U$ simultaneously with $Q_0$,  $U_0$   $i$, $\Omega$, $H$ and $\lambda_0$ as free parameters via a non-linear least squares routine. A Levenburg-Marquardt algorithm was used here through the python module LMFIT \citep{LMFIT}. 

From equation \ref{eq:H}, we can see that $\tau_4$ introduces a phase shift, $\lambda_0$. However, in the case of axial symmetry in the orbital plane about the WR, $\tau_4=0$ and consequently $\lambda_0=0$. Since we are uncertain about the asymmetries in the plane, we will consider two fitting scenarios: 
\begin{enumerate}
\item  Case A :  $\tau_1=\tau_2=0$; $\tau_3\ne \tau_4 \ne 0$ ($\lambda_0 \ne 0$)  
\item  Case B :  $\tau_1=\tau_2=\tau_4=0$; $\tau_3 \ne 0$ ($\lambda_0=0$)
\end{enumerate}
Case B will be further subdivided in Ba and Bb where the inclination anlge will be fixed at either $18^{\circ}$ or $67^{\circ}$, respectively. 

In case A, the values of the fitted parameters yield similar results to \cite{Drissen} (see Table \ref{tab:pol}) with a new  inclination of $\sim 69^\circ$ as opposed to the previous $\sim 67^\circ$. This slight difference is likely due to the new ephemeris and period adopted for phase folding the polarimetric data. The only caveat in this case is the justification for the phase shift relative to phase $\phi=0$ (WR in front). Additional free electrons originating from a colliding-wind shock cone may be responsible. Consequently, a small positive phase shift may be expected in the presence of a strong aberration effect involving the wind-expansion and orbital speeds twisting the shape of the shock cone. While some authors consider this effect to be negligible \citep{Moffat1998}, it has been shown to exist and needed to explain some observations \citep{Lomax}.  However, as seen below in section  \ref{Cw}, the aberration angle in the modelled wind-wind collision zone does in fact appear to be negligible. This is indicative that the case B fits may be more appropriate for WR 148.

 \begin{figure}
     \centering
     \includegraphics[scale=0.4]{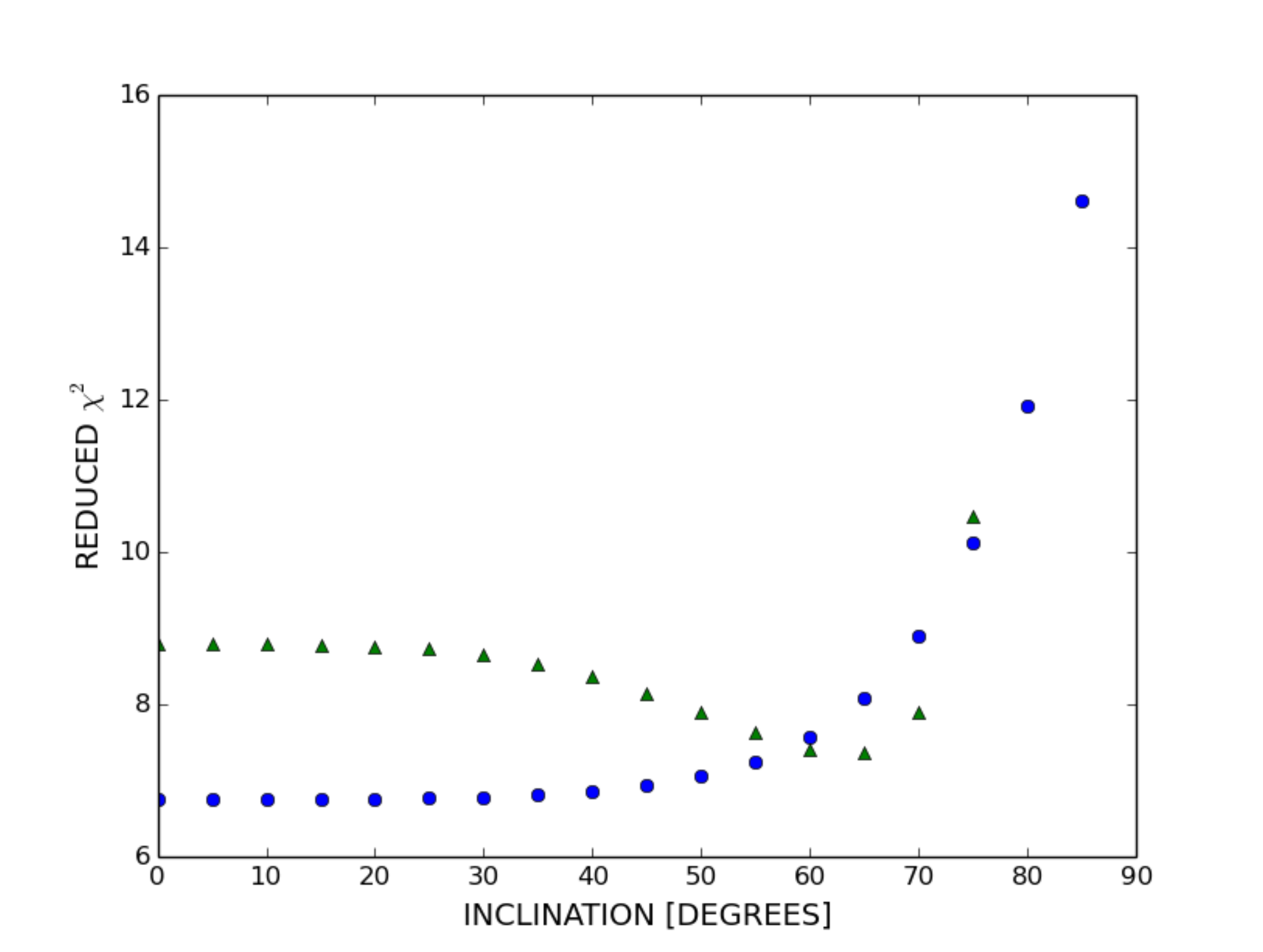}
     \caption{Reduced $\chi^2$ as a function of orbital inclination for the polarimetric model. The green triangles correspond to case A and the blue circles to case B. }
     \label{fig:chipol}
 \end{figure}
 
 \begin{table} 
 \begin{center}
   \caption{Best-Fitting values for the BME polarization model}
   \begin{tabular}{ l c c c  }
     \hline \hline
      Parameter &  Case A &  Case Ba & Case Bb \\ \hline 
     $i$ (deg) &  69.4 $\pm$ 1.9 & 18 (fixed) & 67 (fixed)     \\
     $Q_0$(\%) &  -0.58 $\pm$ 0.02  & -0.773 $\pm$ 0.014 & -0.58 $\pm$ 0.02  \\
     $U_0$(\%) & -0.70 $\pm$ 0.02 & -0.535 $\pm$ 0.014 & -0.51 $\pm$ 0.02 \\
     $\Omega$ (deg)&  145.1 $\pm$ 3.6 & 196.9 $\pm$ 3.5 &  190.8 $\pm$ 4.0  \\
     $H$ (\%) & 0.17 $\pm$ 0.04  & 0.128 $\pm$ 0.008 & 0.231 $\pm$ 0.016   \\
     $\lambda_0$ (deg) & 26 $\pm$ 5  & 0 (fixed) & 0 (fixed)\\
         \hline
   \end{tabular}
   \label{tab:pol}
 \end{center}
 \end{table}

Regarding case B, the fitting procedure is arguably more difficult. This is because of the loss of a free parameter, $\lambda_0$, resulting in a fit with unrealistically large errors, thus rendering the values of the optimized parameters meaningless. Instead, we proceed by fitting four parameters and fixing the last one, inclination. Fixed inclination values range from 0$^\circ$ to 90$^\circ$ with an increment of 5$^\circ$. At each inclination angle, the reduced $\chi^2$ is calculated for the observed Q and U values to the models in equation \ref{eq:QU2} as a means to compare with the other fits. For consistency, the reduced $\chi^2$ was also computed for case A. The results are illustrated in Fig. \ref{fig:chipol}. 

According to the reduced $\chi^2$, it appears that, for case B, the goodness of the fits increases as $i$ decreases. This is true until $\sim50^\circ$, below which, the fits are seemingly identical. That being said, there is a clear global minimum in the reduced $\chi^2$ for case A. The question remains whether the dip is significant or not. Considering that the reduced $\chi^2$ are rather far from 1, the best value for an ideal fit, no case formally fits well. This may be caused by a combination of underestimated observational errors \citep{Bastien} or intrinsic variability of the stellar wind (e.g. due to clumpy structures).

\begin{figure} 
    \centering
    \includegraphics[scale=0.4]{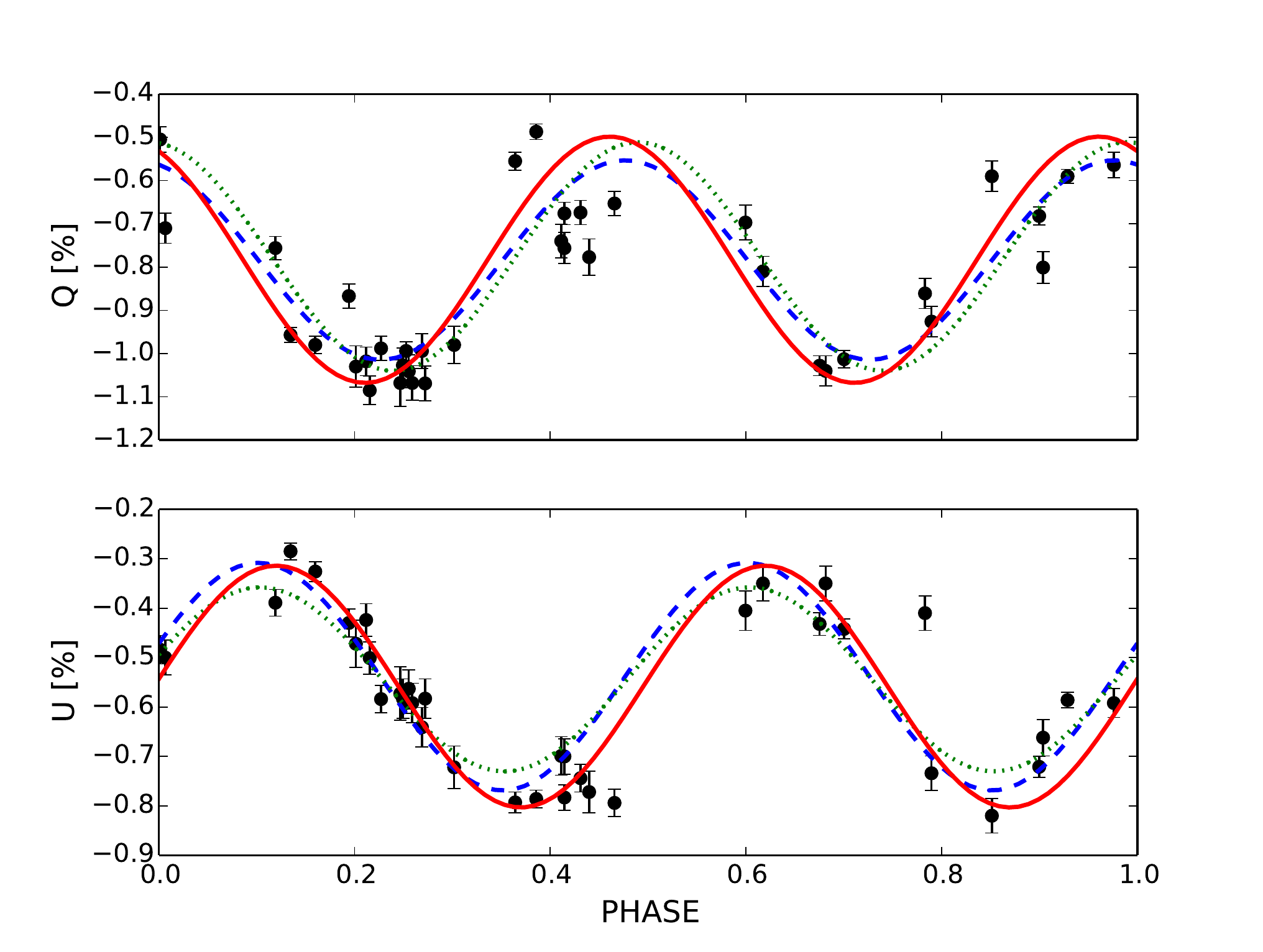}
    \caption{Stokes' parameters $Q$ and $U$ plotted as a function of orbital phase. The red solid curve represents the fit at 69.4$^\circ$ (case A), the blue dashed curve at 18$^\circ$ (case Ba) and the green dotted curve at 67$^\circ$ (case Bb). The polarimetric observations are provided from \citet{Drissen}.}
    \label{fig:QUvsPh}
\end{figure}

\begin{figure}
    \centering
    \includegraphics[scale=0.4]{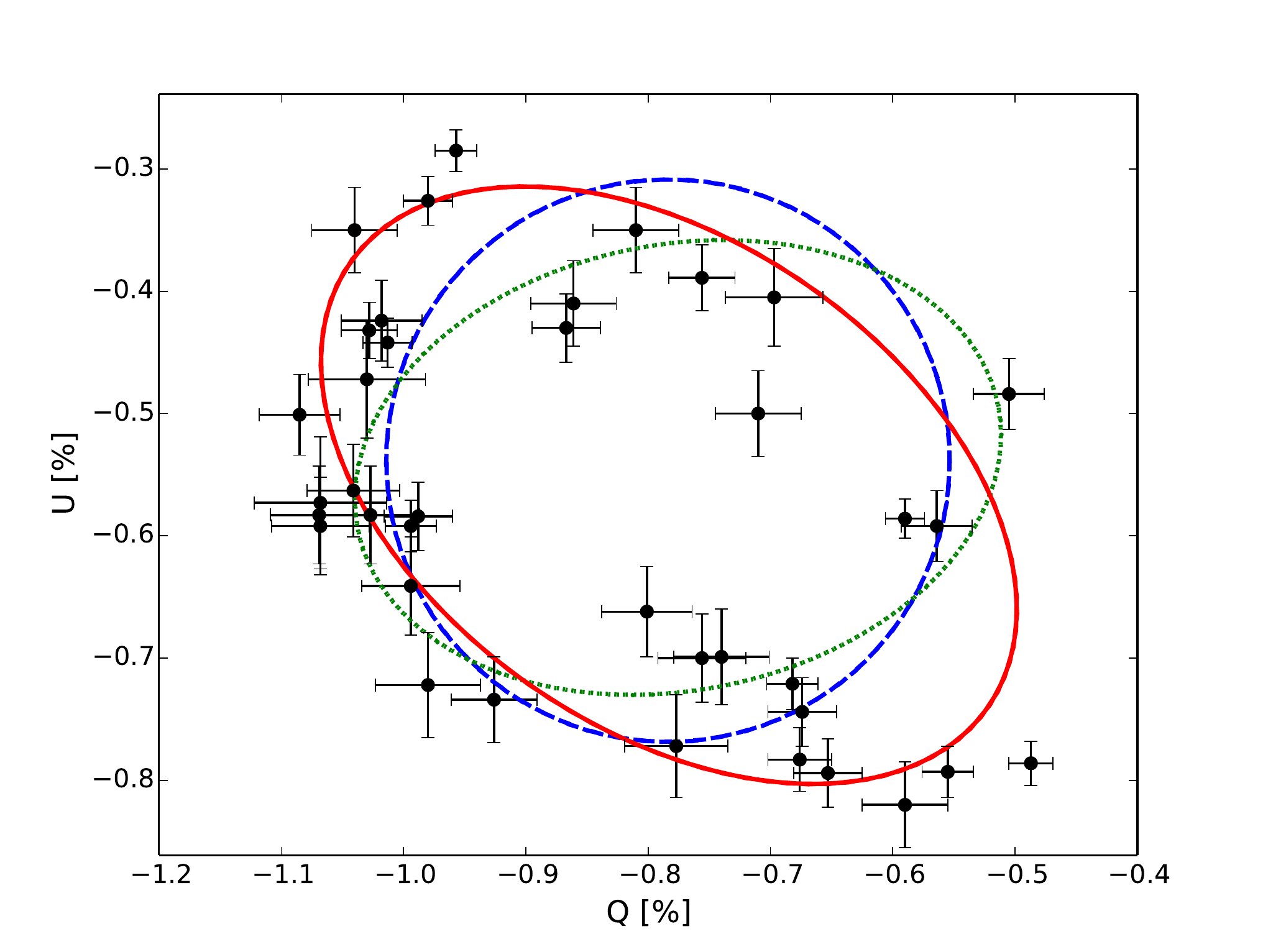}
    \caption{Polarimetric variations plotted in the $Q-U$ plane. The red solid curve represents the fit at 69.4$^\circ$ (case A), the blue dashed curve at 18$^\circ$ (case Ba) and the green dotted curve at 67$^\circ$ (case Bb). Overplotted is a typical single-point 2$\sigma$ error bar.}
    \label{fig:QU}
\end{figure}

Fig. \ref{fig:QUvsPh} shows the fitted $Q$ and $U$ curves as a function of orbital phase at the optimized angle of $69^\circ$ for case A and at both the newly and previously determined angles of  $18^\circ$ and $67^\circ$ for cases Ba and Bb, respectively. No fit appears to be strongly favoured. However,  the same cannot be said for the fits in the Q-U plane (see Fig. \ref{fig:QU}). Although the scatter is quite large, the ellipse of the observed $Q$ vs $U$ data points appears to be quite eccentric. This is most exactly reproduced with case A, implying that the free electron density in the plane is indeed asymmetric.  On the other hand, in the case B produced $Q$ vs $U$ ellipses, at $18^\circ$ it is hardly even eccentric and, even worse, at $67^\circ$ the ellipse appears to be tilted the wrong way. Still, perhaps the scatter is intrinsic to the WR-star (not related to the binary motion) and coincidently the few trailed-off points take the shape of an ellipse. If this is the case then the circular fit at $18^\circ$ may very well best describe the variations of the Stokes' parameters in the $Q-U$ plane.

Regardless of the most likely case, estimation of the confidence intervals for inclinations determined polarimetrically is imperative. According to \cite{error} a quantifiable measure of the data's quality is given by the factor
\begin{equation}
\gamma= \left(\frac{A}{\sigma} \right)^2 \frac{N}{2},
\end{equation}
where $N$ is the number of observations, $\sigma$ is the mean error and $A$ is the amplitude of the polarimetric variability defined by 
\begin{equation}
A=\frac{|Q_\text{max} - Q_\text{min}|+|U_\text{max} - U_\text{min}|}{4}.
\end{equation}
Provided that the polarimetric data have 38 observations each with a standard deviation of $\sigma=0.035\,\%$ and an amplitude of $A=0.23\,\%$, we obtain a $\gamma$ factor of $\simeq750$. It is stated that the polarimetric bias towards higher angles decreases with increasing $i$ and $\gamma$. In other words, one needs better polarimetric precision and/or many more observations to determine accurately lower inclination angles.

Referring to \cite{error}'s Fig. 1 indicates that even with a given precision of $\gamma=1200$, if the true orbital inclination is 10$^\circ$, one can then obtain a best-fitted polarimetric angle ranging from 30 to 60$^\circ$. This window only widens as $\gamma$ decreases as is the case for the WR 148 observations. A denser data set with an enhanced statistical precision is therefore needed to provide a significant constraint.

\subsection{Revisiting the photometry}
\label{phot}
In \citet{Marchenko}, a compilation of light curves was fitted using a modified version of the \citet{Lamontagne} atmospheric eclipse model. The original proposed model (which we use here) is a geometrical paradigm to characterize the phase-dependent scattered light of the O-star as it passes through the WR wind. It assumes a spherically symmetric WR wind and a negligible O-star wind where Thomson scattering by the free electrons in the WR wind is the dominant process. We will assume that the WR wind obeys a velocity expansion law with $\beta=1$ which is probably not unusual for a large, O-like WNLh star. Under these circumstances, \citet{Lamontagne} derived the following analytical expression for the light curve:
\begin{equation} \label{eq:phot1}
\begin{split}
\Delta m = \Delta m_0 + \mathcal{A} \left[  \frac{2}{\sqrt{(1-\epsilon^2)(1-b^2)}} \left[ \arctan \left(\frac{1+b}{1-b}\right) + \right. \right. \\
\left. \left. 
\arctan \left( \frac{1+b}{1-b} \tan \left( \frac{\arcsin \epsilon}{2} \right) \right)  \right]  \right],
\end{split}
\end{equation}
where $\Delta m_0$ is a constant, $\epsilon = \sin i \cos 2 \pi \phi$ and $\mathcal{A}$ is the amplitude defined by
\begin{equation} \label{eq:phot2}
\mathcal{A}= \frac{ (2.5 \log \euler )k }{(1+ I_{WR}/I_O)},
\end{equation}
$k$ and $b$ are given by
\begin{equation} \label{eq:phot3}
k = \frac{\alpha \sigma_t  \Mdot}{4 \pi m_p \varv_{WR}^\infty a},
\end{equation}
\begin{equation} \label{eq:phot4}
b = \frac{R_\star/a}{\sqrt{1- \epsilon^2} },
\end{equation}
where $\sigma_t$ is the free electron Thomson cross section, $m_p$ is the proton mass, $\varv_{WR}^\infty$ is the terminal velocity of the WR-star, $a$ is the orbital separation, $I_{WR}/I_O$ is the WR- to O-star light ratio at the observed wavelength, $\Mdot$ is the WR mass-loss rate, $\alpha$ is the number of free electron per baryon and $R_\star$ is the WR-star core radius. Obtaining the best-fitted $\mathcal{A}$ parameter holds the potential of determining other fundamental parameters, although coupled, such as the light ratio or the mass-loss rate.

Fitting equation \ref{eq:phot1} to the observations requires an $R_\star/a$ ratio. The separation, $a = 7.5 (1 + 1/q)/\sin i$ $\,\Rsol$, varies from $17 - 46 \,\Rsol$  for $i=67^{\circ}$ to $18^{\circ}$, respectively. The photospheric radius is albeit more ambiguous to determine. For massive H-rich WRs, $R_\star$ can range anywhere from 10 to 25 $\Rsol$ \citep{WN}.  To avoid scenarios where the photospheric radius is larger than the separation, we will assume a constant fraction of $R_\star/a=0.5$. In any case, the final result does not depend sensitively on the ratio.

\begin{figure}
    \centering
    \includegraphics[scale=0.4]{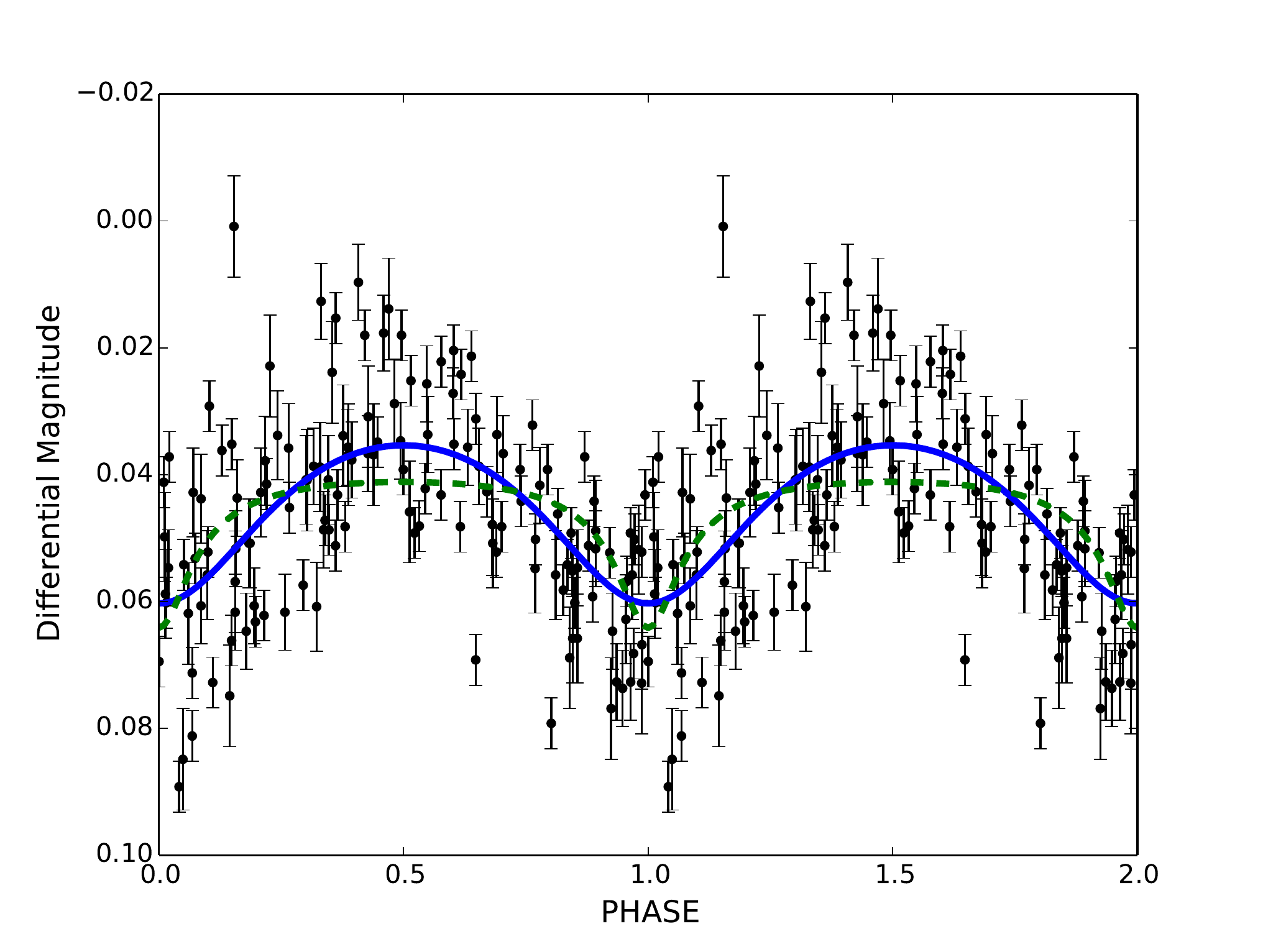}
    \caption{Phased light curve. Photometric observations are provided from \citet{Bracher}, \citet{Antokhin}, \citet{Moffat1986} and \citet{Marchenko}. The blue solid curve represents the fit at 18$^\circ$ and the green dashed at 67$^\circ$. 2$\sigma$ error bars are overplotted. }
    \label{fig:phot}
\end{figure}

\begin{figure}
    \centering
    \includegraphics[scale=0.4]{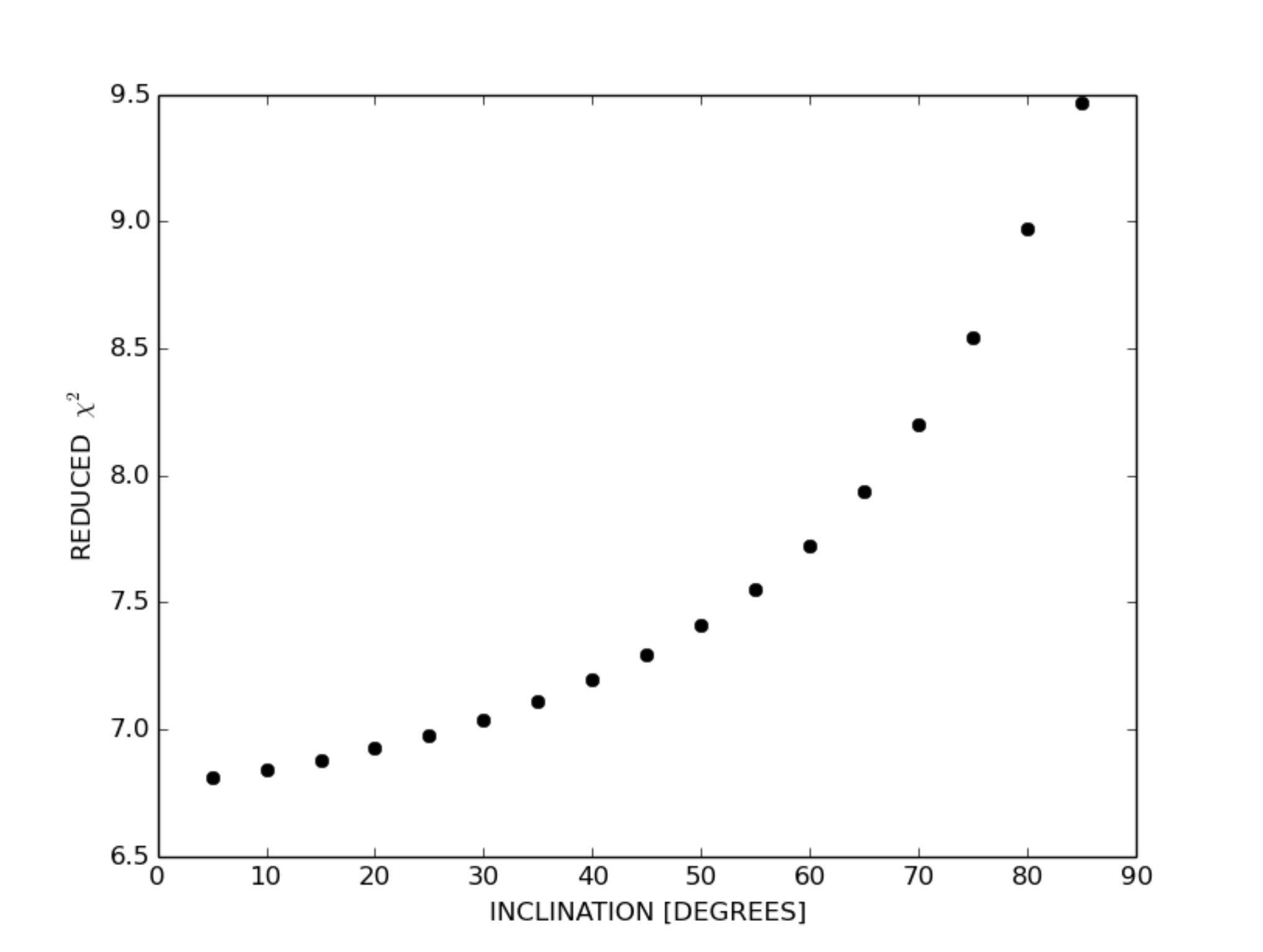}
    \caption{Reduced $\chi^2$ as a function of orbital inclination for the photometric model}
    \label{fig:chiphot}
\end{figure}

Presented in Fig. \ref{fig:phot} is the phased light curve provided from all previous photometric observations in the V band: \citet{Bracher}, \citet{Antokhin}, \citet{Moffat1986} and \citet{Marchenko}. Despite the large scatter (which may be intrinsic to the WR star and possibly due to inhomogeneities in the WR wind), we notice a dip at 0.0 phase. This arises as the O-star passes behind the WR and more of its light is Thomsom scattered out of the line of sight. Biased by the presumed orbital inclination of 67$^\circ$, \citet{Marchenko} fixed $i$ to this value and fit the phased light curves to a modified version of equation \ref{eq:phot1} with as free parameters the amplitude, $\mathcal{A}$, and zero point, $\Delta m_0$, along with enhanced ionization effects from the companion. Here we apply the orignal atmospheric eclipse model from \citet{Lamontagne} so we ignore the companion's extended light source. Ideally, though, the inclination angle should be left as a free parameter. However, with all three parameters left free in the \cite{Lamontagne} atmospheric eclipse model, the fits were unsatisfactory as the errors of the fitted paramers were larger than the values themselves. This is due to the degeneracy between $\mathcal{A}$ and $i$ in equation \ref{eq:phot1} at very low inclinations.

Similar treatment to the polarization was therefore used: ranging $i$ from 0$^\circ$ to 90$^\circ$ and fitting the phased light curves.  At increments of 5$^\circ$, the reduced $\chi^2$ was also computed. We find that the observational data fit best with the model at low inclinations. Indeed, Fig. \ref{fig:chiphot} shows a decrease in the reduced $\chi^2$ towards smaller inclination. However, no global minimum is achieved. Again, there is no convergence in the reduced $\chi^2$ to 1.0, probably for the same reasons as the polarimetry, i.e. due to clumping of the WR wind.

For comparison's sake, two synthetic light curves, obtained while fixing the orbital inclination at $18^\circ$ and $67^\circ$, are shown in Fig. \ref{fig:phot}. The fitted parameters are indicated in Table \ref{tab:phot}. The atmospheric eclipse model coincides best with the observations at lower inclinations. We will henceforth reject the previous $67^\circ$ polarimetric inclination and only consider the newly derived $18^\circ$ inclination.

\begin{table}
\begin{center}
   \caption{Best-fitting values for the \citet{Lamontagne} atmospheric eclipse model}
  \begin{tabular}{ l c c c  }
    \hline \hline
     Parameter &  $i=18^\circ$ & $i=67^\circ$  \\ \hline 
    $\Delta m_0$ [mag] &  0.477 $\pm$ 0.006 &  0.649 $\pm$ 0.006   \\
    $\mathcal{A}$ &  0.017 $\pm$ 0.002 &  0.0030 $\pm$ 0.0005   \\
        \hline
  \end{tabular}
  \label{tab:phot}
\end{center}
\end{table}

\subsection{Colliding Winds} 
\label{Cw}
Like in most massive binaries, where both stars have substantial stellar winds, we expect to see evidence of colliding winds (CWs). The shock region where the winds collide can produce considerable X-ray emission. Similarly, this phenomenon is often manifested as extra line emission in some spectral lines. 
Thermal X-rays have already been found in WR 148 from \citet{XMM}'s analysis of XMM-Newton observations. Here we explore the excess line emission arising from the wind-wind collision zone.

\begin{figure}
    \centering
    \includegraphics[scale=0.4]{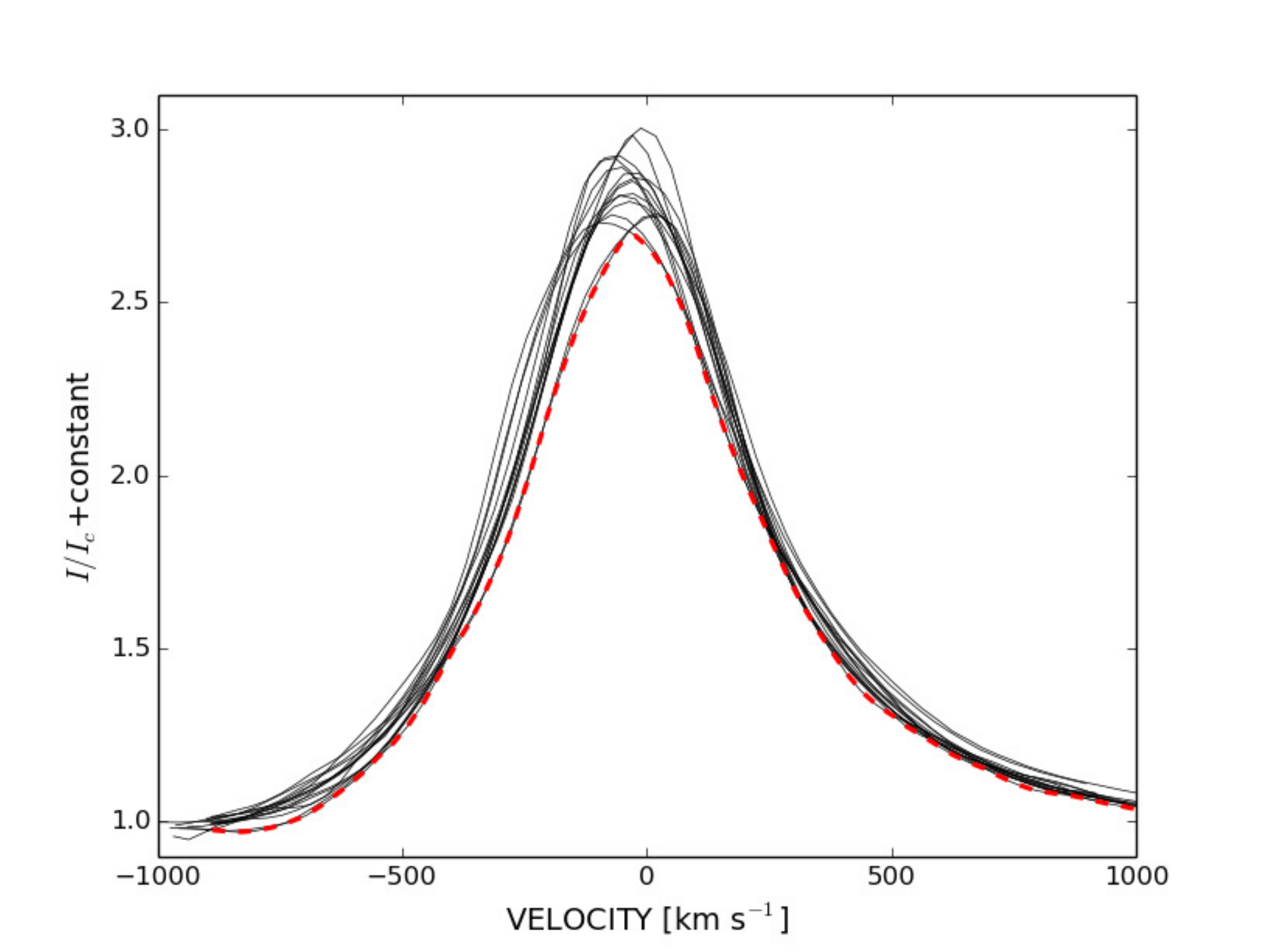}
    \caption{Line profiles of He {\sc ii} $\lambda$4686 shifted to the WR frame. The red dotted curve denotes the synthetic unperturbed base profile.}
    \label{fig:minspec}
\end{figure}

Of particular interest are the line-profile variations of He {\sc ii} $\lambda$4686 because this line is highly sensitive and susceptible to CW \citep[e.g.][]{MarchenkoCW}. In order to extract the excess emission of He {\sc ii} $\lambda$4686, we first need to construct an underlying template of this line, which is then subtracted from each single spectrum in the WR frame and then returned to the observer's frame. Any residual flux is then considered to be excess emission, assuming the underlying profile is constant around the orbit. The base profile is obtained by taking the minimum flux across the line profile, shifted to the WR frame at each pixel (see Fig. \ref{fig:minspec}).

\begin{figure}
    \centering
    \includegraphics[scale=0.4,trim={1.0cm 3cm 1.0cm 3cm},clip]{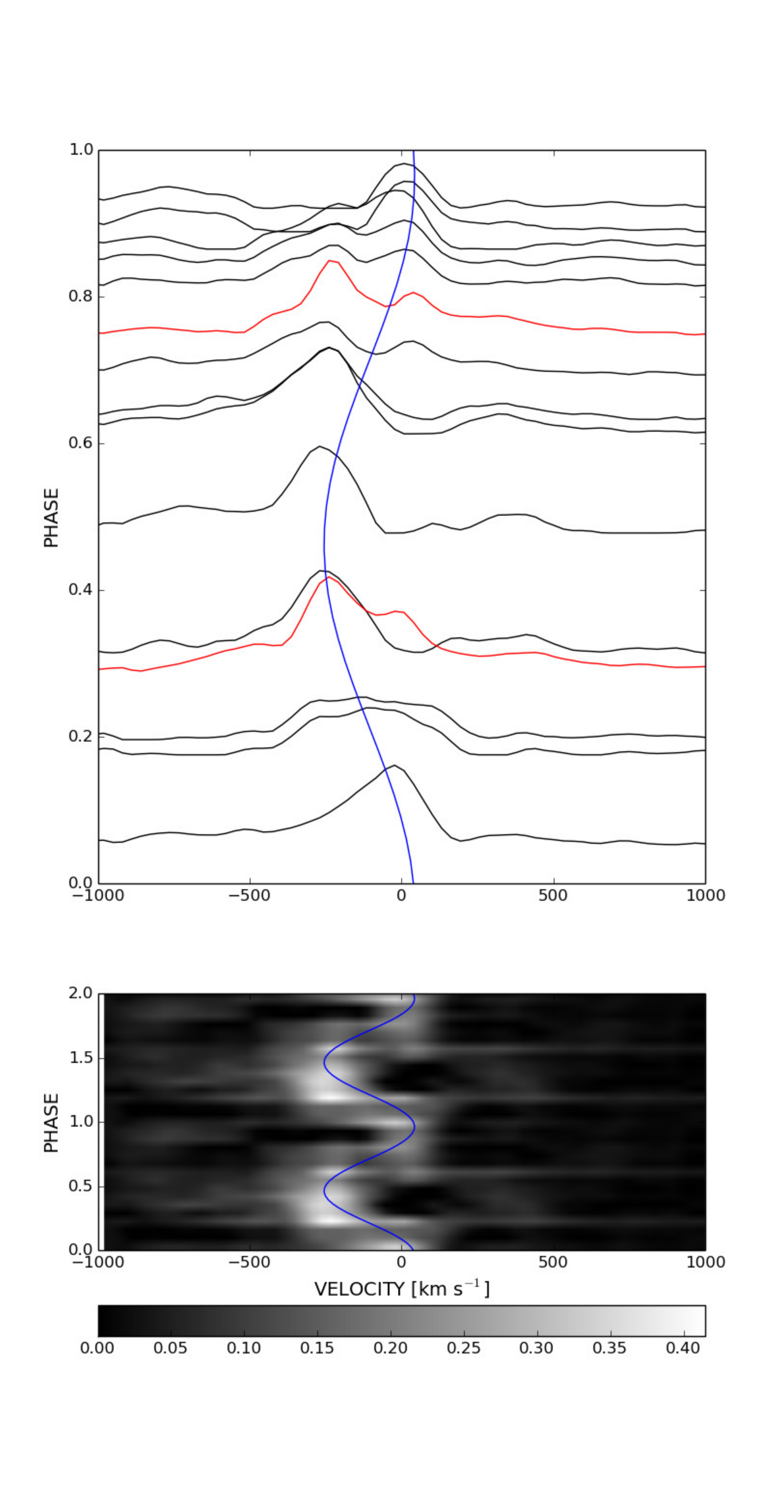}
    \caption{Excess emission in He {\sc ii} $\lambda$4686 varying with phase along the $x$ axis. The red profiles are from the Keck data and the others from OMM. The blue curves represent the CW fit to the excess emission with parameters listed in Table \ref{tab:cw}.  }
    \label{fig:cw}
\end{figure}

\begin{figure}
    \centering
    \includegraphics[scale=0.4]{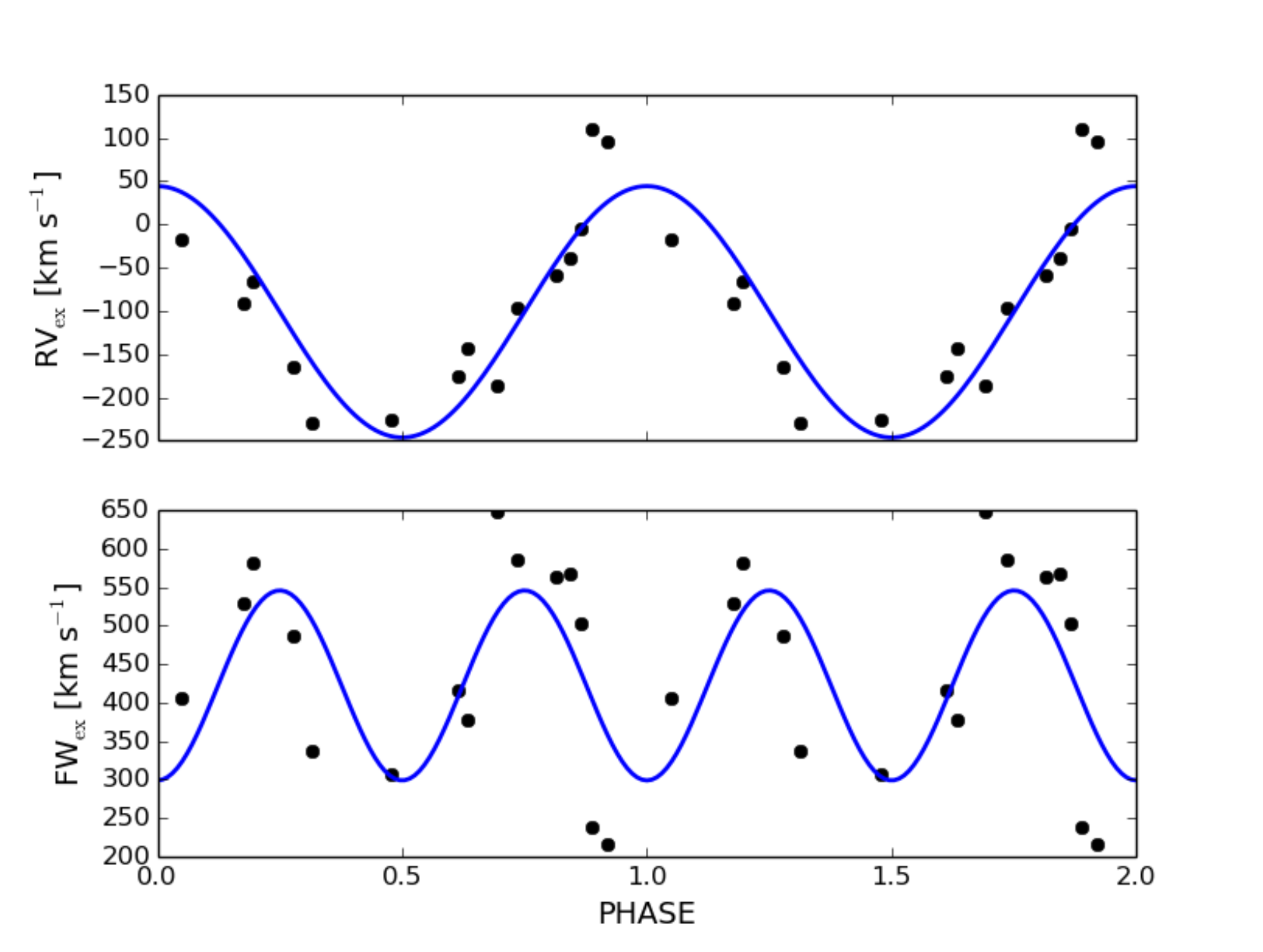}
    \caption{ RV$_\text{ex}$ and FWHM$_\text{ex}$ plotted  as a function of orbital phase. The blue solid curve represents the fit at 18$^\circ$. }
    \label{fig:cwfit}
\end{figure}

The excess emission occurs from the shock-heated material viewed at different angles as the O-star orbits the WR-star. Fig. \ref{fig:cw} demonstrates this phase-dependent behaviour of the extra emission. \citet{Luhrs} first attempted to model these periodic modulations based upon the geometry of the wind-wind collision zone. This was further developed by \citet{Hill}. Adopting the formalism from \citet{Hill}, the radial velocity of this excess emission ($\text{RV}_{\text{ex}}$) and full width (${FW}_{\text{ex}}$) can be characterized by 
\begin{equation}
\label{eq:cw}
 \begin{aligned}
 \text{RV}_{\text{ex}}&= c_2 + \varv_{\text{strm}} \cos \theta \sin i \cos 2 \pi ( \phi -\delta\phi), \\
\text{FW}_{\text{ex}}&= c_1 + 2 \varv_{\text{strm}} \sin\theta \sqrt{1-\sin i ^2 [\cos 2 \pi (\phi-\delta\phi)]^2}, 
  \end{aligned}
\end{equation}
where $c_1$ and $c_2$ are constants, $\theta$ is the the half-opening
angle of the cone, $\varv_{\text{strm}}$ is the streaming velocity of the excess-emitting material in the shock cone and $\delta\phi$ is the aberration angle owing to the orbital motion relative to the wind speed. Combining the OMM and Keck data to improve the phase coverage, we measure $\text{FW}_{\text{ex}}$ at one third maximum and $\text{RV}_{\text{ex}}$ from the centroid of the line at this height. Equations \ref{eq:cw} are fitted simultaneously to the observations while fixing the inclination angle at $18^{\circ}$. The results from the fitting procedures are presented in Table \ref{tab:cw} and shown in Fig. \ref{fig:cwfit}. We note that the aberration angle is effectively zero, which further supports the case B polarimetry scenario (see section \ref{pol}).

On one hand, using the \citet{Usov} formula we can deduce the wind momentum ratio, $\eta$, from only the value of the half opening angle, $\theta$:
\begin{equation} \label{eq:Usov}
\theta (\text{rad}) \simeq 2.1 \left( 1- \frac{\eta^{2/5}}{4}\right) \eta ^{1/3},
\end{equation}
for $10^{-4} \le  \eta \le 1$, with
\begin{equation} \label{eq:eta}
\eta \equiv \frac{\dot{M}_O \varv_{O}^\infty}{\dot{M}_{WR} \varv_{WR}^\infty},
\end{equation}
where $\varv_{O}^\infty$ and $\varv_{WR}^\infty$ are, respectively, the terminal velocities of the O- and WR-stars. With a half opening-angle of $\theta=78.6 \pm 4.0^\circ$, we find a rather large momentum ratio of $\eta=0.51 \pm 0.13$. 
We take note of a similar short period WR+O system with a rather high half-opening angle as well:  V444 Cygni. Indeed, V444 Cygni is a well studied  WN5+O6 colliding-wind binary system ($P=4.212$ d) where $\eta\sim0.44$ can deduced from $\theta=75^\circ$ \citep{Lomax}. This is comparatively high in contrast to the generally obtained values of $\eta<0.05$ from other WR+O colliding wind binaries \citep[e.g.][]{eta1,eta2}. On the other hand, defined by equation \ref{eq:eta}, $\eta$ can also be determined using the typical mass-loss rates and terminal velocity values associated with each star's spectral type. 

The WR's mass loss rate can be estimated polarimetricaly using equation 6 from \citet{StLouis}:
\begin{equation}
\Mdot (\Msol \,\text{yr}^{-1}) = \frac{2.33 \times 10^{-7} A_p a(\Rsol) \varv_\text{WR}^\infty (\text{km s}^{-1}  ) }{(1+\cos^2 i) f_c I},
\end{equation}
where $f_c$ is the fraction of the total light in the filter used coming from the companion O star, $I$ is a dimensionless integral defined in \citet{StLouis} and $A_p$ is the semi-major axis of the ellipse in the $Q-U$ plane, which can otherwise be written as
\begin{equation}
A_p = H (1 + \cos^2 i).
\end{equation}
We determine $A_p (\%)=0.24$ using $H (\%)=0.128$ and $i=18^{\circ}$.

Given the spectral types of each star, $f_c$ is obtained from
\begin{equation}
f_c = \frac{I_O}{I_{WR} + I_{O}}=\left(1+10^{-0.4 \Delta M_v}\right)^{-1},
\end{equation}
where $\Delta M_v = M_v (WR) - M_v(O)$ is the difference between the respective absolute visual magnitudes of the WR- and the O-star. Taking $M_v (WR) = - 7.0$, the mean value between $-6.8$ \citep{Crowther} and $-7.2$ \citep{WN}, and $M_v (O) = -5.21$ \citep{ParamsO} yields $f_c=0.16 \pm 0.04 $ and likewise, $I_{WR}/I_O=5.2\pm 1.0$. Finally, taking  $\varv_\text{WR}^\infty=2000$\,\kms (this paper from inspecting low resolution IUE spectra) and $I$=18.6 (computed numerically with $\beta=1$, $R_\star/a=0.5$ and  $\varepsilon=1$, as defined in \citet{StLouis}) we obtain a polarimetric mass-loss rate of $\log \Mdot_\text{pol} (\,\Msol \yr \, ) =-5.2 \pm 0.3$.

Alternatively, an estimate of the WR's mass-loss rate can be derived photometrically via equations \ref{eq:phot2} and \ref{eq:phot3}. Using  $\alpha=0.5$ (for a wind dominated by He$^{++}$), $I_{WR}/I_O=5.2$, $\varv_{WR}^\infty=2000$\,\kms and the best-fitted amplitude, $\mathcal{A}$, from Table \ref{tab:phot}, we find a photometric mass-loss rate an order of magnitude larger with $\log \Mdot_\text{phot} (\,\Msol \yr \, ) =-4.3 \pm 0.3$  (see Table \ref{tab:mdot}).

\begin{table}
\begin{center}
   \caption{Best-Fitting values for the colliding wind model \citep{Hill}} 
  \begin{tabular}{ l c c  }
    \hline \hline
     Parameter &   $i=18^\circ$  \\ \hline
    $c_1 $ [\kms] & -101 $\pm$ 22		\\
    $c_2 $ [\kms]& -3827 $\pm$ 1400		\\
    $\varv_\text{strm}$ [\kms]&  2230 $\pm$ 700   \\
    $\theta$ [deg] & 77.9 $\pm$ 4.8    \\
    $\delta \phi$ [deg] &  0.0 $\pm$ 5.6   \\
        \hline
  \end{tabular}
  \label{tab:cw}
\end{center}
\end{table}    
   
\begin{table}
\begin{center}
 \caption{Mass loss rates obtained photometrically and polarimetrically}
  \begin{tabular}{ l c c  }
    \hline \hline
     Parameter &  $i=18^\circ$  \\ \hline 
    $\log \Mdot_{\text{phot}}$ [$\Msol$\,yr$^{-1}$] & -4.3 $\pm$ 0.3 \\
    $\log \Mdot_{\text{pol}}$ [$\Msol$\,yr$^{-1}$]& -5.2 $\pm$ 0.3    \\
        \hline
  \end{tabular}
  \label{tab:mdot}
\end{center}
\end{table}

Many factors can contribute to the discrepancy between the photometric and polarimetric mass-loss rates. First, we point out that the photometric mass-loss rates are far more sensitive to changes in the inclination angle than the polarimetric values. In fact, increasing the inclination slightly, from $18^\circ$ to $20^\circ$, decreases $\log \Mdot_\text{phot} (\,\Msol \yr \, )$ to $-4.5 \pm 0.3$,  whereas $\Mdot_\text{pol}$ remains constant. Second, both the polarimetric and photometric mass-loss rates depend on the chosen $\beta$ for the velocity law and the $R_\star/a$ ratio. Even though we have fixed them to their most likely values ($\beta=1$ and $R_\star/a=0.5$), varying them within their uncertainty boundaries can lead to a whole order of magnitude difference on $\Mdot$. Finally, some fundamental assumptions in the models themselves, such as a negligible O-star wind contribution, may not be entirely justified in this case, especially when the wind momentum fraction is rather large. The errors on the mass-loss rates, being computed via propagation of errors, are therefore most likely underestimated.

Due to the uncertainty on the WR's mass loss rate, rather than attempt to re-derive the wind momentum fraction using  equation \ref{eq:eta}, we use the best fitted value for $\eta$ from the CW model to further constrain $\dot{M}_{WR}$. Considering typical values for the terminal velocity and mass-loss rate for an O5V star, $v_O^{\infty}=2900$\,\kms \citep{vinfO} and $\log \dot{M}_O (\,\Msol \yr \, ) =-5.9$\, \citep{vinfO}, yields $\log \dot{M}_{WR}=-5.4\pm0.3$ which is most compatible with the polarimetric mass loss rate.


Up to this point, we have not yet considered the influence of radiative braking. As the WR wind approaches the O-star, the increased O-star radiation can inhibit the WR wind, thus potentially increasing the half-opening angle of the wind-wind interaction cone. This phenomenon has been thoroughly studied by \citet{Braking} and we follow their prescription to determine whether radiative braking has an effect in WR 148 or not. To do so, we first need to evaluate the following parameters: $\hat{d}=d/d_{rb}$ and $\hat{P}=1/(\eta P_{rb})$ where $d_{rb}$ is the solution of 
\begin{equation}
d_{rb} = 1 + \left( \frac{d_{rb}}{\gamma}  \right)^{(1-\alpha)/(1+\alpha)},
\end{equation}
\begin{equation}
P_{rb} = \frac{4 \beta^{\beta} d_{rb}^2}{(2+\beta)^{2+\beta}},
\end{equation}
and
\begin{equation}
\gamma = \frac{4}{3} \left( \frac{L_O}{L_{WR}} \right)^2 \frac{2 G M_{WR}}{(v_{WR}^{\infty})^2 R_O},
\end{equation}
where $L_O$ and $L_{WR}$ are, respectively, the luminosities of the O- and the WR-star, $R_O$ is the O-star's radius, $d=a/R_O$, $\alpha=0.5$ is the CAK line-distribution exponent, $\beta=1$ is the velocity law index and $G$ is the Gravitational constant. Radiative braking becomes important if $\hat{d} > 1$ and $\hat{P} > \hat{d}^2$. Using $\log L_O (L_{\astrosun})=5.21$ and $R_O=11\,\Rsol$, appropriate for an O5V star \citep{ParamsO} and $\log L_{WR} (L_{\astrosun})=6.2$ \citep{WN}, we find $\hat{d}=0.24$ and $\hat{P}=0.04$. Radiative braking therefore appears to be insignificant in this case.

\subsection{Runaway status}
Runaways are stars with abnormally high peculiar space velocities, typically above 30\,\kms for massive stars \citep{Dray}. As a result, they can be found high above the Galactic plane if the star is ejected in an appropriate direction. Two scenarios are generally accepted for producing massive-star runaways: 
\begin{enumerate}
\item  Binary ejection scenario (BES) :  In a binary system, a supernova explosion (SN) may disrupt the system and eject the optical component. Surviving bound systems, though rare, also receive a recoil \citep{BES}. \\
\item  Dynamical ejection scenario (DES) : In dense forming open clusters, runaways arise from close gravitational encounters between stars \citep{DES}.
\end{enumerate}
An alternative theory is expulsion of the binary component in a hierarchical triple system upon the SN explosion of the tertiary.

WR 148's runaways status was highly suspected since \citet{Moffat1979}. However, little is known on its ejection mechanism.  Since the companion for WR 148 is now confirmed to be non-compact, the BES hypothesis is no longer a viable option, leaving dynamical or tertiary ejection mechanisms as the only possible scenarios.  

Taking the Hipparcos proper motions, the spectroscopic radial velocity and estimated distance\footnote{Galactic WR stars have been newly calibrated based on revised near-infrared absolute magnitudes in \citet{Rosslowe}. Since WR 148's distance is more sensitive to the WR-star's absolute visual magnitude uncertainty than on the contribution of the O-star companion, we adopt the distance estimate from \citet{Rosslowe}. } (see Table \ref{tab:hip}), we recreate WR 148's space velocity and retrace its trajectory back in time. This was done in an attempt to determine its origin. As done in \citet{Hoogerwerf}, the orbits are integrated numerically with a fourth-order Runge-Kutta method in a Galactic potential described with a three component model: a logarithmic potential for the halo and a Miyamoto-Nagai potential for the disk and bulge of the Galaxy with parameters recently refined by \citet{galpot}. Prior to the orbit integration, the star's heliocentric velocity, $v_{\star}$, must be corrected for the velocity of the Sun relative to the local standard of rest (LSR), $v_{\astrosun}$, and for the rotational velocity of the LSR relative to the Galactic center, $v_\text{LSR}$. In the Galactic barycentric frame, $v_{\star}$ will thus be converted to
\begin{equation}
v_\text{gal}=v_{\star}+v_{\astrosun}+v_\text{LSR},
\end{equation}
where  $v_{\astrosun}=(10.00,\,5.23,\,7.17)$\,\kms \citep{vSUN} and $v_\text{LSR}=(0.0,\,220.0,\,0.0)$\,\kms \citep{vLSR}. 
Because of the large errors on the proper motions and distance, we conducted 1000 Monte Carlo simulations to scope the parameter error space. We obtained an average travel time (from its current location back to the Galactic plane) of 5.1$\pm$2.0\,Myr, an averaged peculiar velocity of $202\pm$51\,\kms  at its curent location and an averaged peculiar velocity of $204\pm$44\,\kms at its past location leaving the plane.  The errors given here are statistical $1\sigma$ errorbars. Though less likely, results in the $2-3\sigma$ range are conceivable. 

The trajectory time is close to the upper limit of massive WR-stars lifetimes. However, it is certainly in the realm of possibility, especially within the $2\sigma$ level. During this ballistic trajectory, we find that WR 148's peculiar velocity remains roughly constant. \citet{Moffat1998} also computed the peculiar tangential velocity (in direction of Galactic longitude and latitude) for several WR and O star runaways using the Hipparcos proper motions. Taking into account the peculiar radial velocity, we confirm that their obtained tangential peculiar velocity of $194\pm110$\kms for WR 148, is consistent with our result. However, despite utilising essentially the same available Hipparcos data,  our uncertainty on this value is roughly halved. The reason for this improvement can be explained from the use of the more recently revised distance estimate for WR 148. \citet{Moffat1998} assumed a 30\% uncertainty on the distances, yielding $7.72\pm2.32$kpc, while we adopted a distance of 7.05$\pm$1.16kpc from \citet{Crowther}. The distance being the main contribution to the uncertainty for the tangential velocity, this deacrease on the error essentially halves the errorbar.  Also, we note that this velocity is very similar to the high observed peculiar velocity of the WN8h runaway WR 124, close to 200\,\kms \citep{WR124}.

If WR 148 was ejected via a DES, we should be able to retrace its steps back to its parent OB association.  However, this test was inconclusive because not only is the uncertainty on the system's spatial parameters high, but the localizations of the known OB associations \citep{OBasso} are mostly limited to the solar neighborhood and are not so well known near the vicinity where the ejection starting point lies in the Galactic plane. 

Though DES runaways favour single runaways, there is still a non negligible binary fraction. According to the N-body simulations (four-body encounters) performed by \citet{Leonard}, 10\,\% of the ejected runaways are predicted to be binaries. These binaries typically have high mass ratios above 0.5 and eccentricities between 0.4 and 0.7, although it is mentioned that for short period binaries, the eccentricity can decrease rapidly due to tidal circularisation. These properties are all found in WR 148. An extensive study of N-body simulations in massive clusters (three body scatterings) was followed up by \citet{Perets}. They concluded that dynamically ejected binaries become increasingly rare at higher ejection velocities and essentially no binaries are expected to be ejected with velocities above 150\,\kms. With a peculiar space velocity of $\sim202\pm51$\,\kms, WR 148 is definitely a unique case among the fastest massive runaway stars. Within the $\sigma$ range, WR 148's peculiar space velocity is just achievable with the maximum predicted velocity for DES runaways.

Perhaps the system was expulsed by means of the proposed tertiary method. To evaluate the likelihood of this scenario we first need to determine the geometry of a hierarchical triple system. This configuration remains stable if the separation of the outer binary is significantly wider than the inner binary's separation \citep{triple1}. This translates to, 
\begin{equation}
\alpha = \frac{a_2}{a_1} \ll 1,
\end{equation}
where $a_1$ and $a_2$ are respectively the semi-major axes of the outer and inner orbit, regardless of their eccentricities. Determining $\alpha$ analytically can only be done in special cases \citep[see][]{triple2}. Therefore, obtaining a stricter stability criterion is beyond the scope of this paper. In the following, we will assume that $\alpha=0.1$ will suffice to ensure stability.

If the SN explosion of the tertiary decouples the outer orbit, the inner system will be ejected with its orbital velocity 
\begin{equation} \label{eq:K12}
K_{1,2}=\frac{m_3}{m_1+m_2+m_3} v_\text{orb},
\end{equation}
with
\begin{equation}
v_\text{orb}=\sqrt{\frac{G(m_1+m_2+m_3)}{a_1}},
\end{equation}
where $v_\text{orb}$ is the outer systems orbital velocity, $m_3$ is the mass of the tertiary and $m_1+m_2$ is the mass of the inner orbit. Solving equation \ref{eq:K12} for $m_3$ yields
\begin{equation} \label{eq:m3}
m_3=\frac{\kappa + \sqrt{\kappa^2 - 4 \kappa (m_1+m_2)}}{2}
\end{equation}
where $\kappa=\frac{K_{1,2}^2 a_1}{G}$. This allows us to estimate the tertiary's required mass to eject the inner binary with WR 148's observed peculiar velocity. 

The parameters in equation \ref{eq:m3} need to be known at the moment of WR 148's ejection. We know that the system's present orbital separation is $\sim46\,\Rsol$. Considering a WR mass loss rate of $10^{-5.3} \,\Msol$\,yr$^{-1}$ and a WR phase lifetime of $\sim0.5$ Myr [corresponding to  $\sim10$\% of its descendent O-star lifetime \citep{Meynet} assumed to be equivalent to its travel time, i.e. $\sim5$ Myr] we estimate that the WR-star lost approximately $2.5 \Msol$. Whether this mass loss had been transferred to the companion or not shouldn't significantly alter the binary's configuration. In that case, a system with WR 148's present configuration to be ejected with an initial peculiar of $198$ \kms would require a tertiary mass of $137\Msol$. Though high, it is feasible. This scenario would expect WR component to have had an initial mass of  $\sim35\Msol$. However, hydrogen righ WR-stars such as WR 148 are suspected to evolve from massive $65-110\Msol$ O-stars \citep{Maeder2}. To account for this  discrepancy in initial mass, perhaps a formation of a common envelope (CE) caused the system to spiral in and eject its CE. This could expain WR 148's short orbital period. If this is the case, deriving an initial configuration for WR 148 is subject to too many uncertainties. Though, as ejection through hierarchical triple systems are rather inefficient at producing high-velocity runaway binaries (because the large separation needed for stability would require unrealistically large tertiary mass), we consider the tertiary mechanism to be an unlikely ejection mechanism for WR 148.


Another point to consider is the typical timescales of dynamical or tertiary ejections to occur. DES runaways are effectively ejected very near to the zero-age main sequence, whereas the latter ejections are delayed due to the longevity of the tertiary, which could take a few million years depending on its mass. Since the lifetime of the massive WR is already border-line compared to its travel time, it could not afford to wait another 1 or 2 Myr to be ejected. In short, it is unlikely that WR 148 was ejected from the tertiary mechanism and the DES method is prefered.  

Looking further, WR 148 is a prime example of studying the occurence of a double ejection scenario: a DES to explain the current system followed by a future BES.  As the WR component evolves and undergoes a SN explosion, the system may disrupt, possibly providing a supplementary kick. This may give rise to either two single runaways, if the system disrupts, or a binary runaway, if the system remains bound. Given the modeled Galactic potential, the escape velocity at WR 148's current location is estimated to be $\sim548$ \kms. With a space velocity of 281 \kms and an orbital velocity of $\varv_1 = K_1/\sin i \sim 285 \pm 31$ km/s and $\varv_2=K_2\sin i \sim 256\pm 28$ \kms, the gravitational decoupling of the system may eject one or both components to break free from the Galactic gravitational potential.

\begin{figure}
    \centering
    \includegraphics[scale=0.4]{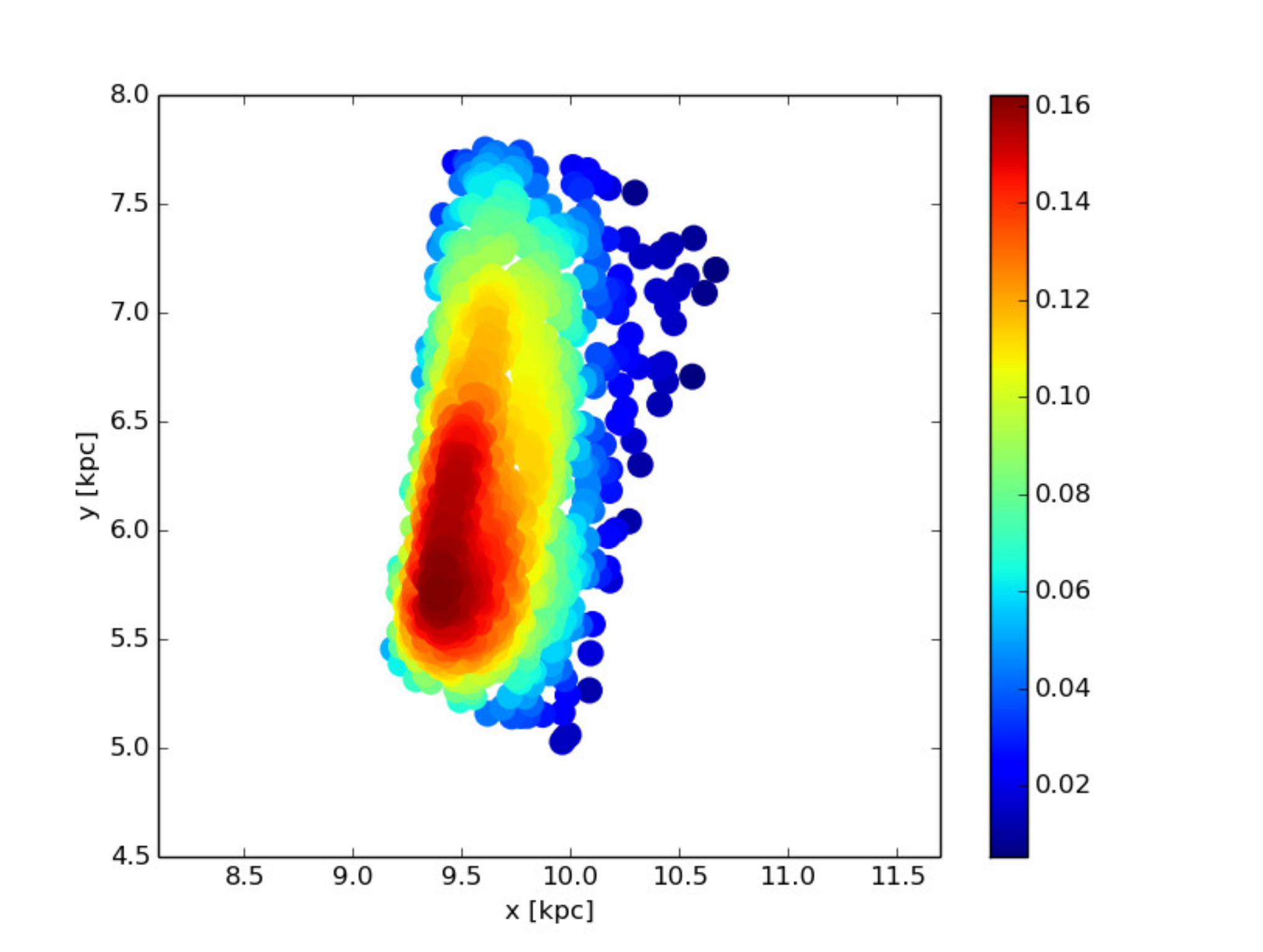}
    \caption{Probability distribution of the possible launching sites of WR 148 back in the Galactic plane. The $x$ and $y$ axis are both centered on the Galactic centre with the $x$ axis passing through the Sun located at $R_0\simeq8.5$kpc perpendicular to the $y$ axis. }
    \label{fig:density}
\end{figure}

\begin{table}
\centering
 \caption{WR 148 spatial parameters}
  \begin{tabular}{ l c c}
    \hline \hline
     Parameters   & Values & Reference \\ \hline 
    $l$ [deg] &  90.0812 & [1] \\
    $b$ [deg] & +06.4694 & [1] \\
    $\mu_{\alpha} \cos \delta$  [mas yr$^{-1}$] & -7.63 $\pm$ 1.39& [1] \\
    $\mu_{\delta}$ [mas yr$^{-1}$]	 & -1.78 $\pm$ 1.13 & [1] \\
    $d$ [kpc] 	& 7.05 $\pm$ 1.16  & [2] \\
    $v_r$ [\kms] & -120.1 $\pm$ 2.3   & [3] \\
    \hline
  \end{tabular}
  \\
   ${\,}^1$\citet{HIP2},${\,}^2$\citet{Rosslowe},${\,}^3$ this paper
 \label{tab:hip}
\end{table}

\subsection{Rotation}
It is instructive to constrain the projected rotation velocity $\varv \sin i$ of the companions, which can yield vital information regarding binary interaction in the system, e.g.\ synchronization and spin-up due to mass transfer. For this purpose, we calculate synthetic spectra for the WR- and O-companions using the Potsdam Wolf-Rayet (PoWR) model atmosphere code (see \citealt{1} and \citealt{2} for more details). The O-star model is calculated with parameters corresponding to an O5 dwarf based on \citet{3}. The WR model is calculated with parameters adopted from \citet{10}, although some adjustments were made to provide a better fit to the specific features which are used here to constrain $\varv \sin i$. 

To simulate the effect of rotation on the spectrum of the O component, we convolve the synthetic O-star spectrum with a rotation profile (virtually an ellipse, \citealt{7}). However, additional broadening that is entangled with the rotation should also be included before $\varv \sin i$ can be derived. These include microturbulence, macroturbulence, and pressure broadening. Pressure broadening is intrinsically accounted for by the PoWR code. The photospheric microturbulence $\xi_\text{ph}$ is set to the typical value of 20\,\kms \citep[e.g.][]{4, 5} and is included in the calculation of the synthetic spectrum. Finally, the macroturbulence $\varv_\text{mac}$, which is typically found to be of the order of 20-90\,\kms \citep[e.g.][]{6}, is treated as a free parameter, and is accounted for by convolving the synthetic spectrum with a corresponding radial-tangential profile \citep{7}.

\begin{figure}
    \centering
    \includegraphics[scale=0.4]{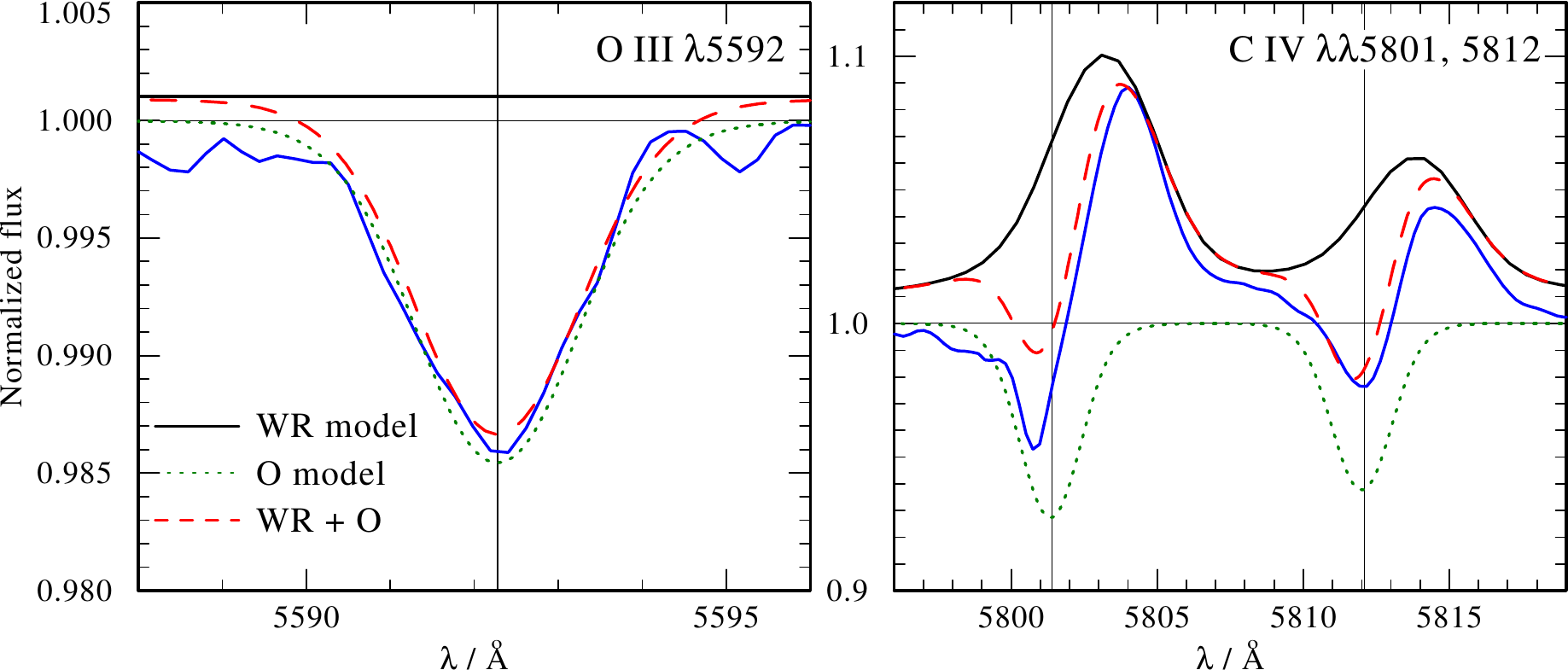}
    \caption{A comparison between the observed spectrum of WR 148 (blue solid line) and the modeled composite spectrum (red dashed line) of the O\,{\sc iii}\,$\lambda 5592$ line (left panel) and the C\,{\sc iv} doublet (right panel). The individual synthetic components are shown in a black solid line for the WR-star and green dotted line for the O-star. The strong absorption feature at $\sim$ 5800 \AA is a residual DIB contamination after attempting to remove the DIB in section \ref{orbit}.}
    \label{fig:OFIG}
\end{figure}

We use the isolated O\,{\sc iii} $\lambda 5592$ line to determine $\varv \sin i$ and $\varv_\text{mac}$ via a 2D $\chi^2$ minimization algorithm, with $0 \le v_\text{mac} \le 90\,$\kms. We obtain $\varv \sin i = 60_{-10}^{+20}$\,\kms and $\varv_\text{mac} = 80_{-30}^{+10}$\,\kms for the O component.  In Fig. \ref{fig:OFIG}, we compare the composite WR + O synthetic spectrum to the 2014 Keck observations, accounting for the secondary's rotation and turbulence. The synthetic O spectrum was convolved with the determined $\varv \sin i$ and $v_\text{mac}$ values. To confirm our results, we also perform an independent analysis of the O\,{\sc iii} feature using the IACOB-BROAD tool, which solves for $\varv \sin i$ in Fourier space \citep{8}, and find $\varv \sin i = 70 \pm 14 \,$\kms. Hence, both methods agree within the errors.

\begin{figure}
    \centering
    \includegraphics[scale=0.4]{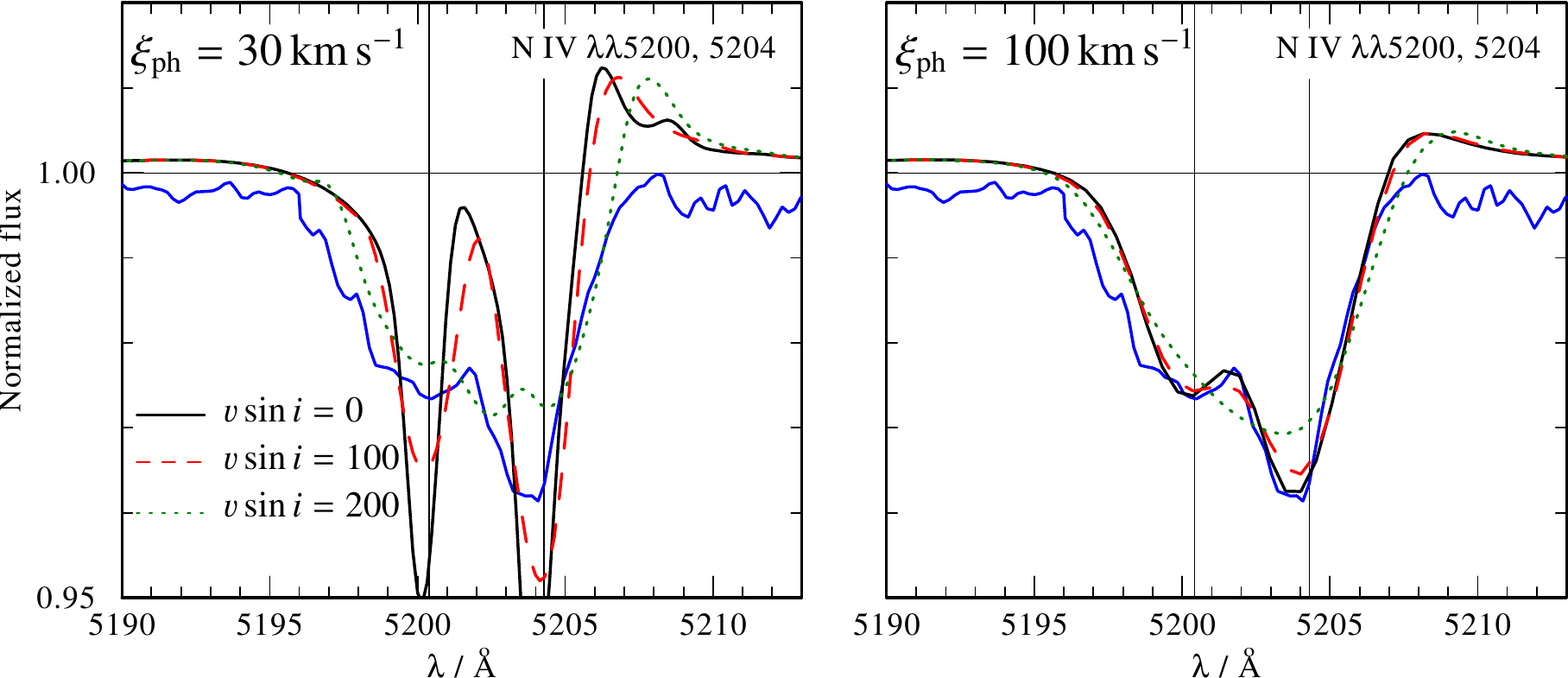}
    \caption{A comparison between the observed spectrum of WR 148 (blue solid line) and three synthetic WR spectrum, assuming $\varv \sin i = 0$ (black solid line), $\varv \sin i = 100$ (dashed red line) and $\varv \sin i = 200$ (dashed red line) of the N\,{\sc iv}\,$\lambda \lambda 5200, 5204$ doublet for $\xi_\text{ph} = 30\,$\kms (left panel) and $\xi_\text{ph} = 100\,$\kms (right panel). }
    \label{fig:WRFIG}
\end{figure}

WR-stars are notoriously known for lacking pure photospheric features, making it difficult to measure their rotation directly. However, careful examination of the Keck spectra revealed a feature which clearly originates in the WR-star and is almost purely photospheric, namely, the N\,{\sc iv}\,$\lambda \lambda 5200, 5204$ doublet. The observation, shown as a blue solid line in Fig. \ref{fig:WRFIG}, shows relatively narrow absorption lines which follow the orbital motion of the WR component. The model calculated for the WR-star confirms that these lines are almost purely photospheric. This means that the observation provides us access to the poorly-understood photosphere at the base of the wind of a WR-star, which is indeed a rare occasion. 

However, determining $\varv \sin i$ for the WR component is significantly more complex than for the O-star component. Firstly, the photospheric microturbulence of late-type WR-stars was poorly studied empirically, let alone their macroturbulence. Secondly, due to the extended WR photospheres, rotation cannot be treated via convolution, but rather a 3D integration algorithm of the intensities is necessary \citep{9}. Because the problem is significantly more complex, here we only give a rough estimate for the possible rotation of the WR component. A thorough and detailed study of the photospheres of this and other late-type WR-stars will be given elsewhere (Shenar et al.\ in prep.). 

For simplicity, $v_\text{mac} = 0$\,\kms is assumed. Then, assuming $\xi_\text{ph} = 30\,$\kms, we find $\varv \sin i \sim 150\,$\kms, although the fit quality is rather poor. Assuming on the other extreme that $\xi_\text{ph} = 100\,$\kms, we can only give an upper limit of $\varv \sin i \lesssim 150$\,\kms. Fig. \ref{fig:WRFIG} illustrates this: The observations (blue solid line) of the N\,{\sc iv} doublet are compared to the synthetic WR spectrum calculated with $\xi_\text{ph} = 30\,$ (left panel) and 100\,\kms (right panel) and with different rotation velocities, as given in the legend. Indeed, this simple analysis implies that large turbulence velocities are present already in the photospheres of WR-stars.

To summarize, we find $\varv \sin i = 60_{-10}^{+20}$\kms for the O component and $\varv \sin i \lesssim 150$\,\kms for the WR component. Assuming that the rotation axes are aligned with the orbital axis and that $i = 18^\circ$, this yields $v_\text{eq} = 194\,$\kms and $v_\text{eq} \lesssim 485\,$\kms for the equatorial velocities of the O and WR components, respectively. 

The measured spin rates of the O-star component in WR+O binaries reveal that $v_\text{eq}$ can range anywhere from 140 to 496\kms \citep{Shara} and the system may simply be synchronized. Adopting $R_\text{O} = 11\,R_\odot$ for the O component as given by \citet{3} for an O5V star, we find $P_\text{rot, O} = 2.9_{-0.8}^{+0.6}\,$d, which is smaller than the measured binary period of $P_\text{orb} = 4.3173\,$d. The WR rotation constraint does not help us further here. However, accounting for the uncertainties on the radius and inclination angle, we cannot forego the idea of synchronization. For instance, using $\varv \sin i=50$\kms, $R_\text{O}=12\Rsol$ and $i=21^{\circ}$, we obtain $P=4.35$ d. Thus, while the system has most definitely circularized, it may in fact be synchronized as well. Though, due to accumulated error propagation, we cannot confirm this with certainty.

\section{Conclusions}
\label{Conclusion}
We summarize our findings briefly as follows:

\begin{itemize}
  \item WR 148 is found to be a normal, massive, close WR+O binary system: the primary is a  H-burning WN7ha star and the secondary is an O5V star. 
  \item This confirms the colliding wind binary scenario, rejecting once again the WR+cc scenario proposed in the past.
  \item Orbital solution is refined: P = 4.317336 d and time for phase zero (WR in front at inferior conjunction) of E = 2 444 825.04 HJD. 
  \item We obtain a mass ratio of $1.1\pm0.1$. Assuming a mass of 37 $\Msol$ for the O star, the WR component has a mass of 33 $\Msol$ and the system has an orbital inclination of $18\pm4{^\circ}$.  
  \item Regarding the previously determined inclination angle of 67${^\circ}$, we re-examine past polarimetric and photometric observations. Via a more appropriate error assessment, the polarimetric results are at best inconclusive requiring better data. The light curve is also now found to behave normally for an atmospheric eclipse of the O-star as it orbits in the WR wind with a low inclinations.  
  \item We deduce a O/WR wind momentum ratio of 0$.51\pm0.13$ from analyzing the excess emission arising from CWs. Adopting typical mass loss rates and terminal velocities for an O5V star, we obtain for the WR component $\log \Mdot (\,\Msol \yr \, )=-5.4\pm0.3$. This is consistent with the mass loss rate derived from polarimetry.  
  \item Runaway status is confirmed. Most likely ejected via dynamical interactions, WR 148 is an extreme runaway with a current peculiar velocity of $\sim197$\,\kms.
  \item WR 148 is currently $\sim800$\,pc from the Galactic plane. It took $\sim5$\,Myr to reach this position starting from the plane, which is marginally acceptable for massive-star lifetimes.  
  \item The runaway's space velocity is not enough to allow for WR 148 to escape the Galactic potential; however, after the SN explosion of the current primary, two single massive runaways could result, with one or both able to escape the Galaxy.
  \item We find a projected rotational velocity of $\varv \sin i = 60_{-10}^{+20}$\kms for the O star and $\varv \sin i \lesssim 150$\,\kms for the WR star. Adopting $\sin i$ from the orbit, leads to high rotation speeds for both stars. Though the system has definitely circularized, we cannot confirm whether it has synchronized.
\end{itemize}

\section*{Acknowledgements}
AFJM and NSL are grateful for financial aid from NSERC (Canada) and FQRNT(Quebec). We  acknowledge the help and support of colleagues at Universit\'{e} de Montr\'{e}al, Universit\"{a}t Potsdam and W. M. Keck Observatory. NDR acknowledges postdoctoral support by the University of Toledo and by the Helen Luedtke Brooks Endowed Professorship. TR acknowledges support from the Canadian Space Agency grant FAST.

The authors wish to recognize and acknowledge the very significant cultural role and reverence that the summit of Mauna Kea has always had within the indigenous Hawaiian community. We are most fortunate to have the opportunity to conduct observations from this mountain. 



\bibliographystyle{apj}
\bibliography{biblio}






\appendix

\section{Mean Keck spectra of WR 148 at two orbital quadratures}

\begin{figure*}
    \centering
    \includegraphics[scale=0.85]{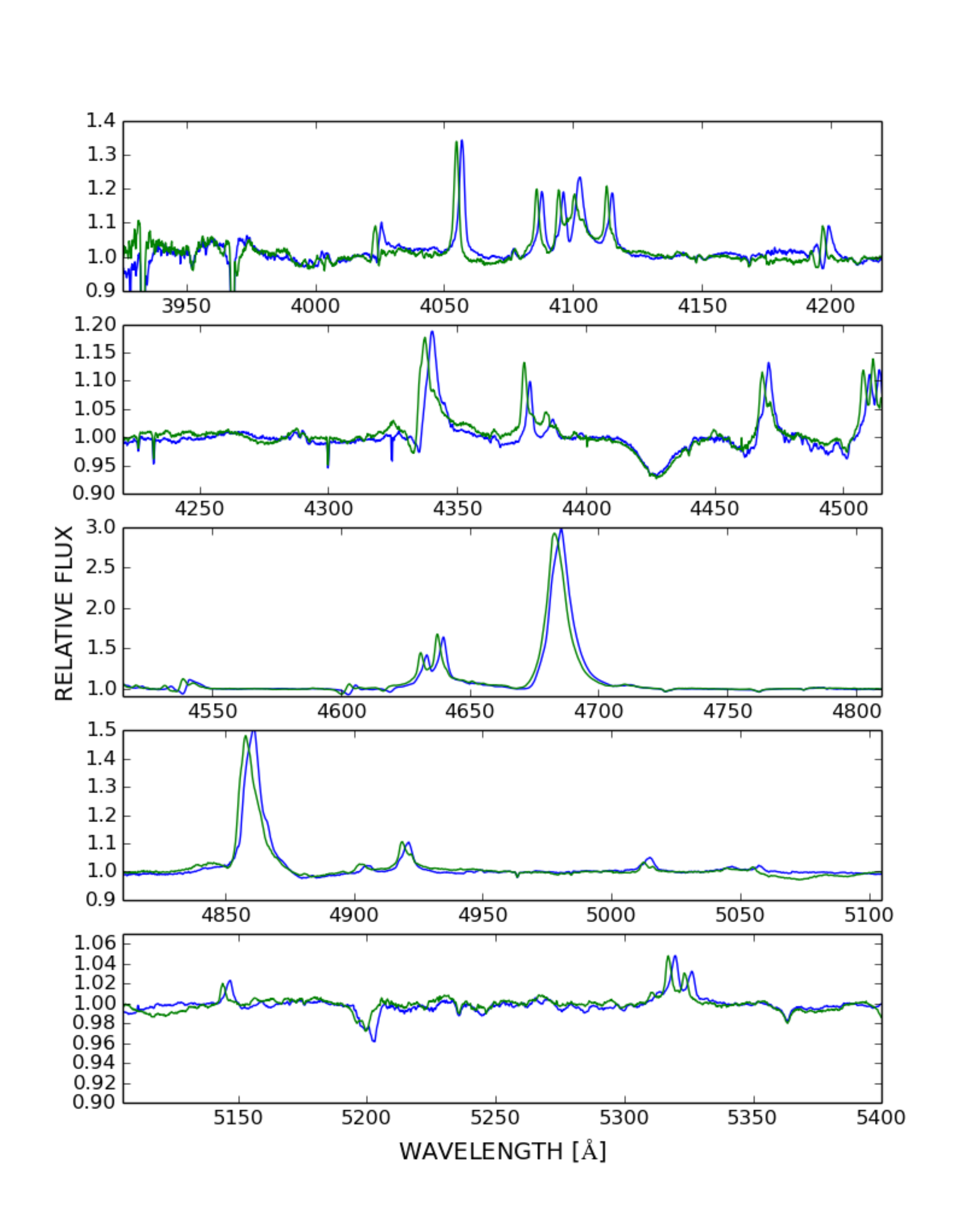}
    \caption{ Mean spectra from the Keck observatory at both quadatures: phase $\simeq 0.279$ in blue and $\simeq 0.736$ in green. }
    \label{fig:esiA}
\end{figure*}

\begin{figure*}
    \centering
    \includegraphics[scale=0.85]{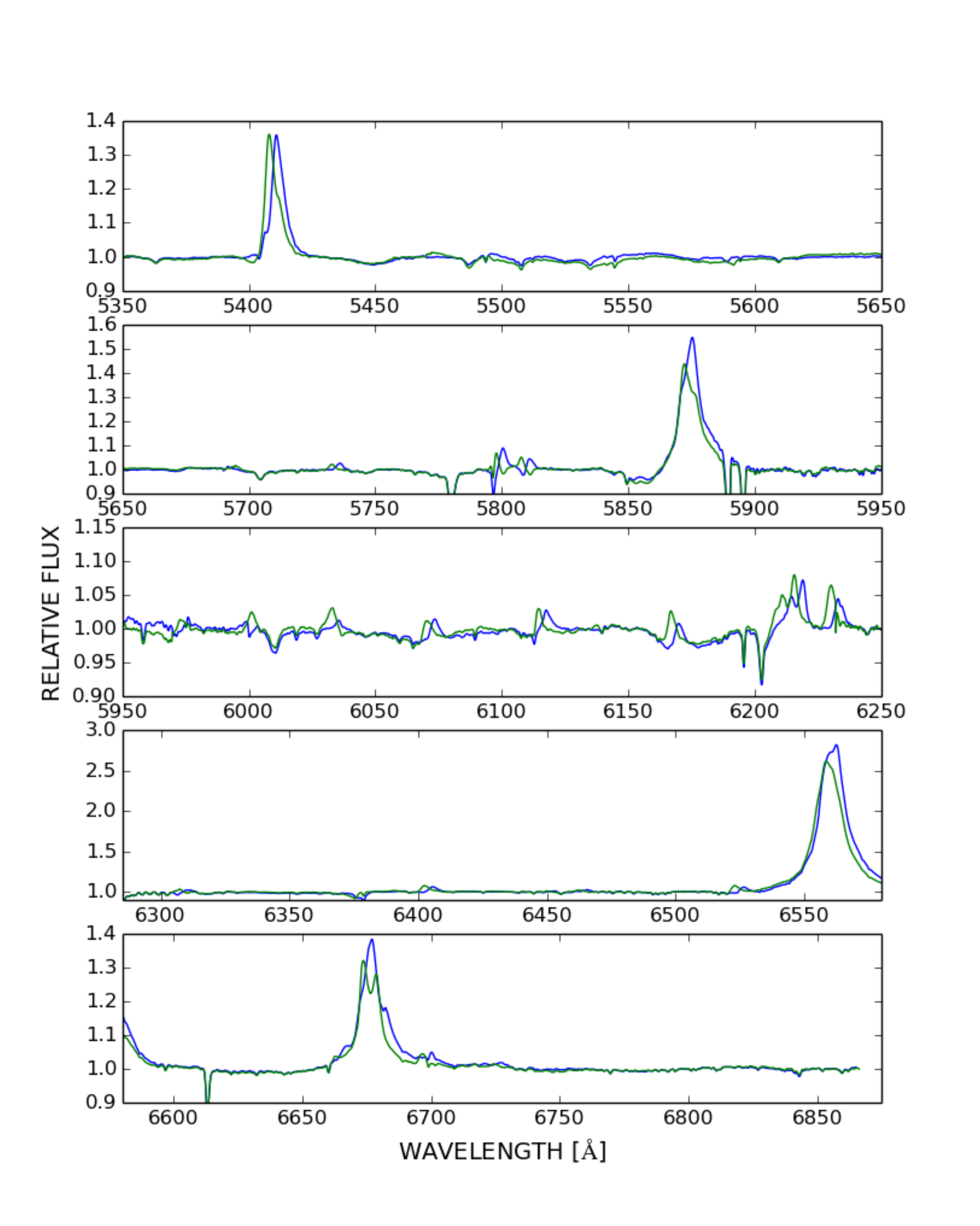}
    \caption{ Continuation of Fig. \ref{fig:esiA} }
    \label{fig:esiB}
\end{figure*}


\bsp	
\label{lastpage}
\end{document}